\documentclass[pra,twocolumn,aps,amssymb,showpacs,superscriptaddress]{revtex4-1}

\usepackage{epsfig}
\usepackage{graphicx}
\usepackage{amsmath} 
\usepackage{amssymb}
\usepackage{enumerate}
\usepackage{hyperref}
\usepackage[normalem]{ulem}
\usepackage{cancel}

\usepackage{color}
\definecolor{rosso}{rgb}{1,0,0}
\definecolor{verde}{rgb}{0,1,0}
\definecolor{blue}{rgb}{0,0,1}
\definecolor{verdescuro}{rgb}{0,0.5,0.5}
\definecolor{rossoscuro}{rgb}{0.7,0.3,0}
\definecolor{bluscuro}{rgb}{0.3,0,0.7}
\definecolor{magenta}{rgb}{1,0,1}

\begin{document}

\title{Spatial emergence of Off-Diagonal Long-Range Order \\ throughout the BCS-BEC crossover}

\author{L. Pisani}
\affiliation{School of Science and Technology, Physics Division, Universit\`{a} di Camerino, 62032 Camerino (MC), Italy}
\author{P. Pieri}
\affiliation{Dipartimento di Fisica e Astronomia, Universit\`a di Bologna, I-40127 Bologna (BO), Italy}
\affiliation{INFN, Sezione di Bologna, I-40127 Bologna (BO), Italy}
\author{G. Calvanese Strinati}
\email{giancarlo.strinati@unicam.it}
\affiliation{School of Science and Technology, Physics Division, Universit\`{a} di Camerino, 62032 Camerino (MC), Italy}
\affiliation{INFN, Sezione di Perugia, 06123 Perugia (PG), Italy}
\affiliation{CNR-INO, Istituto Nazionale di Ottica, Sede di Firenze, 50125 (FI), Italy}


\begin{abstract}
In a superfluid system, Off-Diagonal Long-Range Order (ODLRO) is expected to be exhibited in the appropriate reduced density matrices when the relevant particles (either bosons or fermion pairs) are considered to recede \emph{sufficiently} far apart from each other.
This concept is usually exploited to identify the value of the condensate density, without explicit concern, however, on the spatial range over which this asymptotic condition can effectively be achieved.  
Here, based on a diagrammatic approach that includes beyond-mean-field pairing fluctuations in the broken-symmetry phase at the level of the $t$-matrix also with the inclusion of the Gorkov-Melik-Barkhudarov (GMB) correction, we present a systematic study of the two-particle reduced density matrix for a superfluid fermionic system undergoing the BCS-BEC crossover, when the entities to recede far apart from each other evolve with continuity from largely overlapping Cooper pairs in the BCS limit to dilute composite bosons in the BEC limit.
By this approach, we not only provide the coupling and temperature dependence of the condensate density at the level of our diagrammatic approach which includes the GMB correction, 
but we also obtain the evolution of the spatial dependence of the two-particle reduced density matrix, from a power-law at low temperature to an exponential dependence at high temperature in the superfluid phase, when the inter-particle coupling spans the BCS-BEC crossover.
Our results put limitations on the minimum spatial extent of a finite-size system for which superfluid correlations can effectively be established. 

\end{abstract}

\maketitle

\section{Introduction} 
\label{sec:introduction}
\vspace{-0.2cm}

The concept of Off-Diagonal Long-Range Order (ODLRO) \cite{Yang-1962} is central to superfluid and superconducting systems in the broken-symmetry phase.
It entails the appearance of an asymptotic correlation in the off-diagonal matrix elements in the spatial representation of the one-particle (for bosons) and two-particle (for fermions) reduced density matrices \cite{Yang-1962,DiCastro-Raimondi-2015}.

Accordingly, it could be of interest to assess how this property manifests itself in the context of the BCS-BEC crossover,
whereby the system evolves with continuity from a BCS (fermionic) regime of highly overlapping Cooper pairs to a BEC (bosonic) regime of dilute composite bosons \cite{Physics-Reports-2018}.
In particular, for a Fermi gas this crossover is spanned in terms of the dimensionless coupling parameter $(k_{F} a_{F})^{-1}$ where $k_{F}=(3 \pi^{2} n)^{1/3}$ is the Fermi wave vector with density $n$ and $a_{F}$ the scattering length of the two-fermion problem.
This parameter ranges from $(k_{F}\, a_{F})^{-1} \lesssim -1$ in the weak-coupling (BCS) regime when $a_{F} < 0$, to $(k_{F}\, a_{F})^{-1} \gtrsim +1$ in the strong-coupling (BEC) regime when $a_{F} > 0$, across the unitary limit when $|a_{F}|$ diverges.

In this context, previous theoretical works have specifically considered obtaining only the \emph{condensate density} $n_{0}$, that can be extracted from the values of the above off-diagonal matrix elements \emph{in the asymptotic limit} of large spatial separation.
This has been done for a wide coupling range across the BCS-BEC crossover and for temperatures from zero up to the superfluid critical temperature $T_{c}$,
both at the mean-field level \cite{SMP-2005} and with the inclusion of fluctuation effects at the Gaussian level \cite{Griffin-2007}.
Experimental works with ultra-cold Fermi gases, too, have considered determining the condensate density, mainly at low temperatures around the unitary regime \cite{Jin-2004,Ketterle-2004} and more recently over a wider coupling range \cite{Roati-2020}.
The value of the condensate density at unitarity (both at zero and finire temperature) was also recently obtained by several Quantum Monte Carlo simulations \cite{Astrakharchik-2005,Needs-2010,Mitas-2011,Lee-2020,Alhassid-2020,Bulgac-2020}.                          
In the present work, we complement and extend the above previous theoretical studies about the ODLRO for a Fermi gas undergoing the BCS-BEC crossover, by not only addressing the value of the condensate density as previously done in the literature,
but also determining \emph{over what spatial range this asymptotic correlation is established\/} below the superfluid temperature $T_{c}$ in the broken-symmetry phase.

A characteristic feature of a superfluid is that, owing to broken gauge symmetry, correlations of infinite range establish long-range phase coherence.
This property results into the \emph{static\/} phase-phase (transverse) correlation function to decay as the inverse power of the distance \cite{Foster-1975}.
In the case of the ODLRO of interest here, on the other hand, the two-particle reduced density matrix corresponds to an \emph{equal-time\/} correlation function (cf. Eq.~(\ref{two-particle-reduced-density-matrix}) below and Sec.~\ref{sec:theoretical_approach}-E),
where by equal-time we mean that its calculation implies summing over an infinite set of frequencies (in contrast to a static correlation function for which only the zero frequency is required). 
Through an analysis of its spatial behavior, we will show that, at sufficiently low temperature, this correlation function decays with distance as an \emph{inverse-square\/} law, while only at higher temperatures in the superfluid phase the inverse-proportionality decay 
with distance characteristic of the static phase-phase correlation function is recovered.
This finding is consistent with the progressive irrelevance of the finite-frequency components in the description of a quantum many-body system as the temperature is raised above zero \cite{Hertz-1976}, 
when a crossover temperature from quantum to thermal regimes can be identified.
In the following, this temperature analysis will be carried out in detail with the inter-particle coupling varying along the BCS-BEC crossover.
Our findings are consistent with the occurrence of a \emph{generic scale invariance}, whereby long-range order appears in a whole region of the phase diagram, with power-law decays of correlation functions occurring in an entire phase and not just at an isolated critical point \cite{BKV-2005}.

Notwithstanding the occurrence of the inverse power-law behaviors mentioned above, from the spatial dependence of the two-particle reduced density matrix we shall be able to extract (albeit with some limitations in the temperature interval $0 \le T \le T_{c}$ - see below) a coupling- and temperature-dependent length referred to as the ODLRO length $\xi_{\mathrm{odlro}}$, which turns out to be related to the \emph{inter-pair} (healing) length and, similarly to it, diverges at the critical temperature.
This healing length (sometimes referred to as the ``phase coherence length'' $\xi_{\mathrm{phase}}$) was originally determined as a function of coupling along the BCS-BEC crossover, at zero temperature in Ref.~\cite{PS-1996} as well as a function of temperature in the superfluid phase in Ref.~\cite{PS-2014}, by looking at the large-distance exponential behavior of the static amplitude-amplitude (longitudinal) correlation function.
Here, we will demonstrate that this length scale characteristic of the (massive) static longitudinal correlation function manifests itself also in the (massless) equal-time two-particle reduced density matrix associated with ODLRO.
However, this will turn out to be possible only in the temperature ranges near absolute zero and near the transition temperature $T_{c}$, for reasons related to the fact that determining a time-dependent Ginzburg-Landau equation in the BCS regime was found to be possible only in those temperature regimes \cite{AT-1966}.
Physically, this is because at finite temperature there is the possibility of local conversion of the thermally excited normal excitations to superfluid.
Mathematically, this is due to the intrinsic contribution of the finite-frequency components to the equal-time ODLRO correlation function of interest here.

A distinctive feature of both lengths $\xi_{\mathrm{odlro}}$ and $\xi_{\mathrm{phase}}$ at low temperature is that, as a function of coupling,
they saturate to a minimum value which is of the order of the inter-particle distance.
Recently, this feature has also been highlighted in condensed-matter experiments \cite{Jarillo-Herrero-2021,Deutsher-2021}, where it was used to bring out analogies with the BCS-BEC crossover \cite{Physics-Reports-2018} and, in particular, to identify what would therein correspond to the unitary regime.

We shall further find that an additional length scale (referred to as $\xi_{1}$) enters the ODLRO correlation function, which turns out to be related to the \emph{intra-pair} correlation length (or Cooper pair size) $\xi_{\mathrm{pair}}$ that remains finite at $T_{c}$. 
This length was determined as a function of coupling along the BCS-BEC crossover, at zero temperature in Ref.~\cite{PS-1994} and as a function of temperature in the superfluid phase in Ref.~\cite{PS-2014}.

In practice, achieving these goals will be implemented by relying on the diagrammatic $t$-matrix approximation in the broken-symmetry phase as developed in Refs.~\cite{APS-2003,PPS-2004}, which will enable us to extract the asymptotic spatial behavior 
of the two-particle reduced density matrix of ODLRO with reasonable numerical effort.
In addition, as far as the calculation of the condensate density is concerned, we will go beyond this approach and include also the Gorkov-Melik-Barkhudarov (GMB) correction \cite{GMB-1961} throughout the whole BCS-BEC crossover, which was proved in Ref.~\cite{PPS-2018} to be an important step for a reliable description of the superfluid gap parameter not only in the BCS but even in the unitary regime.

The main results obtained in this article are as follows:  

\noindent
(i) The improved values of the condensate density $n_{0}$, both as a function of coupling and temperature, whose novelty is to
include the GMB correction and which are compared with recent experimental and Quantum Monte Carlo results.

\noindent
(ii) The spatial profiles of the projected density matrix (as obtained by suitably tracing the two-particle reduced density matrix), also
as a function of coupling and temperature, which show visually how the asymptotic value $n_{0}$ is eventually reached for large
separations.

\noindent
(iii) The coupling and temperature dependence of the characteristic length $\xi_{\mathrm{odlro}}$ associated with the ODLRO, whenever it can be extracted from the above spatial profiles and thus compared with the inter-pair healing length $\xi_{\mathrm{phase}}$.

\noindent
(iv) The coupling and temperature dependence of the length $\xi_{1}$ associated with the normal contribution to the two-particle reduced density matrix, which brings out of this density matrix also the other relevant (intra-pair healing) length $\xi_{\mathrm{pair}}$ of the 
BCS-BEC crossover.

\noindent
(v) The coupling and temperature dependence of the distance $R^{*}$, at which the asymptotic correlations entailed by the ODLRO
can effectively (and pragmatically) be reached within a given default uncertainty. 

\noindent
(vi) The effect of the finite size of the system on the apparent value of the condensate system, which can also be extended above the critical temperature.

The article is organized as follows.
Section~\ref{sec:theoretical_approach} sets up the diagrammatic approach to the two-particle reduced density matrix and discusses its various contributions within the $t$-matrix approach.
Section~\ref{sec:numerical_results} reports on the numerical results obtained for the spatial behavior of the two-particle reduced density matrix throughout the BCS-BEC crossover over a wide temperature range, from zero up to (and even above) $T_{c}$.
Section~\ref{sec:conclusions} gives our conclusions.
Additional technical details are given in the Appendices.
Analytic results are given, in Appendix~\ref{sec:appendix-A} for the fermionic one-particle density matrix within the BCS approximation, 
in Appendix~\ref{sec:appendix-B} for the bosonic one-particle density matrix within the Bogoliubov approximation,
and in Appendix~\ref{sec:appendix-C} for the asymptotic behavior of the two-particle reduced density matrix at zero temperature in the BCS limit.


\vspace{-0.4cm}
\section{Theoretical approach} 
\label{sec:theoretical_approach}
\vspace{-0.2cm}

In this Section, the two-particle reduced density matrix for a fermionic system is analyzed in terms of a diagrammatic approach in the broken-symmetry phase.
In this way, not only the condensate density will be identified through the emergence of the asymptotic ODLRO for large spatial separation of fermion pairs, but also the spatial extent for reaching this asymptotic situation will be obtained as a function of coupling and temperature.
To this end, we shall explicitly rely on the $t$-matrix approach developed in Refs.~\cite{APS-2003,PPS-2004} and adapt it to the present circumstances.
In addition, whenever relevant, we shall also include the GMB correction as made to evolve along the BCS-BEC crossover in Ref.~\cite{PPS-2018}.
Throughout, we consider balanced spin populations and set $\hbar = 1$ for convenience.

\vspace{-0.5cm}
\subsection{Diagrammatic approach to \\ the two-particle reduced density matrix}
\label{subsec:general_structure}
\vspace{-0.2cm}

\begin{figure}[t]
\begin{center}
\includegraphics[width=7.4cm,angle=0]{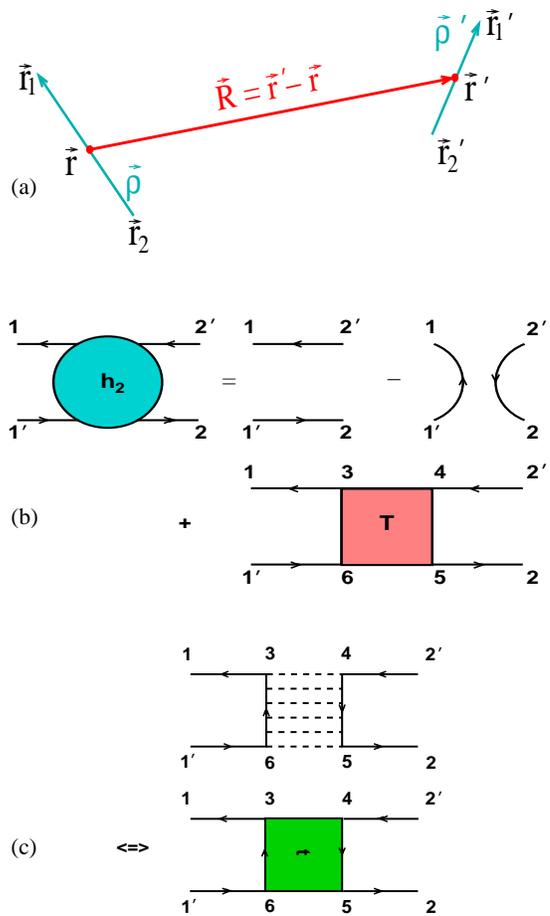}
\caption{(Color online) (a) Spatial coordinates for the two-particle reduced density matrix $h_{2}$, according to the definitions (\ref{grouping-spatial-coordinates}) and (\ref{grouping-spatial-coordinates-primed}).
                                     (b) Diagrammatic representation of $h_{2}$ corresponding to the Bethe-Salpeter equation (\ref{Bethe-Salpeter-equation}),
                                          where the external indices are identified by the dictionary (\ref{compact-notation}) while the internal indices are integrated over.
                                     (c) Pairing fluctuations contribution to $h_{2}$ with the structure of a Maki-Thompson diagram, where the quantity labeled by $t$ stands for a $t$-matrix. 
                                      In panels (b) and (c), full lines represent the single-particle Green's functions (with the arrows pointing from the second to the first of their arguments) and broken lines correspond to the inter-particle interaction.}
\label{Figure-1}
\end{center} 
\end{figure}

The \emph{two-particle reduced density matrix\/} is defined by (cf., e.g., Ref.~\cite{DiCastro-Raimondi-2015})
\begin{equation}
h_{2}(\mathbf{r}_{1},\mathbf{r}_{2};\mathbf{r}_{1'},\mathbf{r}_{2'}) = \langle \psi^{\dagger}_{\uparrow}(\mathbf{r}_{1}) \psi^{\dagger}_{\downarrow}(\mathbf{r}_{2}) \psi_{\downarrow}(\mathbf{r}_{2'}) \psi_{\uparrow}(\mathbf{r}_{1'})\rangle
\label{two-particle-reduced-density-matrix}
\end{equation}
where $\psi_{\sigma}(\mathbf{r})$ is a fermionic field operator with spin $\sigma = (\uparrow,\downarrow)$ and $\langle \cdots \rangle$ stands for an ensemble average.
It will be convenient to group the spatial variables in Eq.~(\ref{two-particle-reduced-density-matrix}) as follows:
\begin{equation}
\hspace{-0.9cm} \left\{ 
\begin{array}{lll}
\!\! \mathbf{r}_{1} = \mathbf{r} + \frac{\boldsymbol{\rho}}{2} \hspace{2.3cm} \boldsymbol{\rho} = \mathbf{r}_{1} - \mathbf{r}_{2} \\
\hspace{2.5cm} \Longrightarrow \\
\!\! \mathbf{r}_{2} = \mathbf{r} - \frac{\boldsymbol{\rho}}{2} \hspace{2.3cm} \mathbf{r} = \frac{1}{2} \! \left( \mathbf{r}_{1} + \mathbf{r}_{2} \right)
\end{array}
\right.
\label{grouping-spatial-coordinates}
\end{equation}
and similarly for the primed quantities
\begin{equation}
\left\{ 
\begin{array}{lll}
\!\! \mathbf{r}_{1'} = \mathbf{r'} + \frac{\boldsymbol{\rho}'}{2} \hspace{2.0cm} \boldsymbol{\rho}' = \mathbf{r}_{1'} - \mathbf{r}_{2'} \\
\hspace{2.5cm} \Longrightarrow \\
\!\! \mathbf{r}_{2'} = \mathbf{r'} - \frac{\boldsymbol{\rho}'}{2} \hspace{2.0cm} \mathbf{r'} = \frac{1}{2} \! \left( \mathbf{r}_{1'} + \mathbf{r}_{2'} \right)
\end{array}
\right.
\label{grouping-spatial-coordinates-primed}
\end{equation}
such that the magnitude of $\mathbf{R} = \mathbf{r}' - \mathbf{r}$ identifies the distance between the center-of-mass coordinates $\mathbf{r}$ and $\mathbf{r}'$ of the two pairs $(\mathbf{r}_{1},\mathbf{r}_{2})$ and $(\mathbf{r}_{1'},\mathbf{r}_{2'})$, respectively, of opposite-spin fermions (cf. \color{red}Fig.~\ref{Figure-1}(a)\color{black}).
For the homogeneous system we are interested in, only three coordinates (say, $\boldsymbol{\rho}$, $\boldsymbol{\rho}'$, and $\mathbf{R}$) suffice to describe the full spatial dependence of $h_{2}$.
In particular, the ODLRO we are after corresponds to the behavior of $h_{2}$ for large values of $R=|\mathbf{R}|$.

When dealing with the broken-symmetry phase, it is convenient to introduce at the outset the Nambu representation \cite{Nambu-1960}, whereby $\Psi_{1}(\mathbf{r}) = \psi_{\uparrow}(\mathbf{r})$ and $\Psi_{2}(\mathbf{r}) = \psi^{\dagger}_{\downarrow}(\mathbf{r})$
with the index $\ell = (1,2)$ distinguishing the two components.
In terms of this representation, the two-particle reduced density matrix (\ref{two-particle-reduced-density-matrix}) can be cast in the form
\begin{eqnarray}
& - & h_{2}(\mathbf{r}_{1},\mathbf{r}_{2};\mathbf{r}_{1'},\mathbf{r}_{2'}) 
\label{two-particle-reduced-density-matrix-compact-form} \\
& = & \langle T_{\tau} [ \Psi_{1}(\mathbf{r}_{1'},\tau) \Psi_{2}(\mathbf{r}_{2},\tau^{++}) \Psi^{\dagger}_{2}(\mathbf{r}_{2'},\tau^{+}) \Psi^{\dagger}_{1}(\mathbf{r}_{1},\tau^{+++})] \rangle
\nonumber 
\end{eqnarray}

\noindent
where $T_{\tau}$ is the imaginary-time order operator and $\tau^{+}$ signify that the imaginary time $\tau$ is augmented by the positive infinitesimal $\eta$.
In addition, with the dictionary 
\begin{equation}
\hspace{-0.4cm} \left\{ 
\begin{array}{lll}
\!\! 1 \longleftrightarrow  \left( \mathbf{r}_{1'},\tau,\ell=1 \right)           \hspace{1.2cm}    2' \longleftrightarrow  \left( \mathbf{r}_{2'},\tau^{+},\ell=2 \right) \\
\\
\!\! 1' \longleftrightarrow  \left( \mathbf{r}_{1},\tau^{+++},\ell=1 \right) \hspace{0.55cm}    2 \longleftrightarrow  \left( \mathbf{r}_{2},\tau^{++},\ell=2 \right)
\end{array}
\right.
\label{compact-notation}
\end{equation}
whose short-hand notation encompasses space, imaginary time, and Nambu indices, $h_{2}$ can be expressed in terms of the two-particle Green's function $\mathcal{G}_{2}$, in the compact form \cite{APS-2003}:
\begin{eqnarray}
&& h_{2}(\mathbf{r}_{1},\mathbf{r}_{2};\mathbf{r}_{1'},\mathbf{r}_{2'}) = - \mathcal{G}_{2}(1,2;1',2')
\nonumber \\
& = & \mathcal{G}(1,2') \, \mathcal{G}(2,1') - \mathcal{G}(1,1') \, \mathcal{G}(2,2') 
\label{Bethe-Salpeter-equation} \\
& + & \int \! d3456 \, \mathcal{G}(1,3) \, \mathcal{G}(6,1') \, T(3,5;6,4) \, \mathcal{G}(4,2') \, \mathcal{G}(2,5) \, .
\nonumber
\end{eqnarray}
Here, $\mathcal{G}$ is the single-particle Green's function and $T$ the many-particle $T$-matrix which formally solves the Bethe-Salpeter equation for $\mathcal{G}_{2}$ (as depicted in \color{red}Fig.~\ref{Figure-1}(b)\color{black}) \cite{APS-2003}.
In any practical calculation based on diagrammatic approaches, suitable approximations have to be selected for $\mathcal{G}$ and $T$.
Their choice in the present context will be discussed below.

The three terms on the right-hand side of Eq.~(\ref{Bethe-Salpeter-equation}) contribute in different ways to the two-particle reduced density matrix and will accordingly be dealt with separately in the following.

\vspace{-0.5cm}
\subsection{Anomalous contribution to $h_{2}$}
\label{subsec:anomalous_contribution}
\vspace{-0.2cm}

The first term on the right-hand side of Eq.~(\ref{Bethe-Salpeter-equation}) does not depend on $\mathbf{R}$ and can be written only in terms of the \emph{anomalous\/} single-particle Green's functions $\mathcal{G}_{12}$ and $\mathcal{G}_{21}$.
With the definitions (\ref{grouping-spatial-coordinates}) and (\ref{grouping-spatial-coordinates-primed}) and the dictionary (\ref{compact-notation}), we obtain for this contribution to $h_{2}$:
\begin{equation}
\mathcal{G}(1,2') \, \mathcal{G}(2,1') = \mathcal{G}_{12}(\boldsymbol{\rho}',0^{-}) \, \mathcal{G}_{21}(\boldsymbol{-\rho},0^{-})
\label{first-term-h_2-general}
\end{equation}
where $\mathcal{G}_{21}(\boldsymbol{-\rho},0^{-}) = \mathcal{G}_{21}(\boldsymbol{\rho},0^{-}) = \mathcal{G}_{12}(\boldsymbol{\rho},0^{-})$ and the negative infinitesimal refers again to the imaginary time domain. 
Physically, the spatial ($\boldsymbol{\rho}$) dependence of $\mathcal{G}_{12}(\boldsymbol{\rho},0^{-})$ accounts for the internal structure of a Cooper pair and identifies the associated pair size $\xi_{\mathrm{pair}}$ at any coupling and temperature \cite{PS-2014}.

In particular, at the mean-field (mf) level one gets:
\begin{equation}
\mathcal{G}_{12}^{\mathrm{mf}}(\boldsymbol{\rho},0^{-}) = \int \!\!\! \frac{d \mathbf{k}}{(2 \pi)^{3}} \, e^{i \mathbf{k} \cdot \boldsymbol{\rho}} \, \frac{\Delta}{2 E(\mathbf{k})} \, \left[ 1 - 2 f(E(\mathbf{k})) \right] \, .
\label{G_12-mean-field} 
\end{equation}
Here, $\Delta$ is the gap parameter (taken real in the following without loss of generality), $E(\mathbf{k}) = \sqrt{ \xi(\mathbf{k})^{2} + \Delta^{2} }$ where $\xi(\mathbf{k}) = \frac{\mathbf{k}^{2}}{2m} - \mu$
with $m$ the fermion mass and $\mu$ the chemical potential, and $f(\epsilon) = \left( e^{\beta \epsilon} +1 \right)^{-1}$ is the Fermi function with inverse temperature $\beta = (k_{B} T)^{-1}$ ($k_{B}$ being the Boltzmann constant).
In addition, in the strong-coupling (BEC) limit whereby $\beta \mu \rightarrow - \infty$, the expression (\ref{G_12-mean-field}) simplifies to the form
\begin{equation}
\mathcal{G}_{12}^{\mathrm{mf}}(\boldsymbol{\rho},0^{-}) \longrightarrow \int \!\!\! \frac{d \mathbf{k}}{(2 \pi)^{3}} \, e^{i \mathbf{k} \cdot \boldsymbol{\rho}} \, \frac{\Delta}{2 \, \xi(\mathbf{k})} = \sqrt{n_{0}} \, \phi (\boldsymbol{\rho})
\label{G_12-mean-field-BEC-limit} 
\end{equation}
since $|\mu| \simeq (2 m a_{F}^{2})^{-1}$ ($\mu < 0$), where
\begin{equation}
n_{0} = \frac{m^{2} a_{F}}{8 \pi} \, \Delta^{2}
\label{condensate-density-mean-field-BEC-limit}
\end{equation}
is the mean-field \emph{condensate density\/} in this limit \cite{SMP-2005} and 
\begin{equation}
\phi (\boldsymbol{\rho}) = \frac{e^{- |\boldsymbol{\rho}|/a_{F}}}{\sqrt{2 \pi a_{F}} \,\, |\boldsymbol{\rho}|}
\label{normalized-two-body-function-BEC-limit}
\end{equation}
the normalized two-fermion wave function in vacuum.

Quite generally, at any coupling and temperature below $T_{c}$, the condensate density can be obtained from the expression \cite{Leggett-2006}
\begin{eqnarray}
n_{0} & = & \int \!\! d\boldsymbol{\rho} \,\, \mathcal{G}_{12}(\boldsymbol{\rho},0^{-}) \, \mathcal{G}_{21}(\boldsymbol{-\rho},0^{-}) = \int \!\! d\boldsymbol{\rho} \,\, \mathcal{G}_{12}(\boldsymbol{\rho},0^{-})^{2} 
\nonumber \\ 
& = & \int \!\!\! \frac{d \mathbf{k}}{(2 \pi)^{3}} \left[ \frac{1}{\beta} \sum_{n} e^{i \omega_{n} \eta} \, \mathcal{G}_{21}(\mathbf{k},\omega_{n}) \right]^{2} \, ,
\label{condensate-density-general}
\end{eqnarray}
where again $\eta$ is a positive infinitesimal and $\omega_{n} = (2n+1)\pi/\beta$ ($n$ integer) is a fermionic Matsubara frequency \cite{FW-1971}.
In particular, at the mean-field level owing to Eq.~(\ref{G_12-mean-field}) the expression (\ref{condensate-density-general}) reduces to the form \cite{SMP-2005}:
\begin{equation}
n_{0} = \int \!\!\! \frac{d \mathbf{k}}{(2 \pi)^{3}} \, \frac{\Delta^{2}}{4 \, E(\mathbf{k})^{2}} \left[ \tanh \left( \frac{\beta E(\mathbf{k})}{2} \right) \right]^{2} \, .
\label{condensate-density-mean-field}
\end{equation}

In Sec.~\ref{subsec:results_anomalous} below, we will compare the results of $n_{0}$ obtained numerically by (six) different levels of approximation:

\noindent
(i) For reference purposes, we will initially consider the mean-field approximation (\ref{condensate-density-mean-field}) to the expression (\ref{condensate-density-general}) \cite{SMP-2005}, whereby the values of the gap parameter $\Delta$ and chemical potential $\mu$ to be inserted 
therein are obtained by solving the coupled
gap and density equations:
\begin{eqnarray}
\hspace{-0.5cm}- \frac{m}{4 \pi a_{F}} & = & \!\! \int \! \frac{d\mathbf{k}}{(2 \pi)^{3}} \left( \frac{1 - 2 f(E(\mathbf{k}))}{2 E(\mathbf{k})} - \frac{m}{\mathbf{k}^{2}} \right) 
\label{BCS-gap-equation} \\
n & = & \!\! \int \! \frac{d\mathbf{k}}{(2 \pi)^{3}} \left( 1 - \frac{\xi(\mathbf{k})}{E(\mathbf{k})} \left[1 - 2 f(E(\mathbf{k})) \right] \right) \, .
\label{BCS-density-equation}
\end{eqnarray}

\noindent
(ii) Inclusion of pairing fluctuations beyond mean field will be first considered at the level of the $t$-matrix approach in the broken-symmetry phase of Ref.~\cite{PPS-2004}.
Accordingly, the gap equation will be kept of the form (\ref{BCS-gap-equation}) while the density equation will be replaced by
\begin{equation}
n = \int \!\!\! \frac{d \mathbf{k}}{(2 \pi)^{3}} \, \frac{2}{\beta} \sum_{n} e^{i \omega_{n} \eta} \, \mathcal{G}_{11}^{\mathrm{pf}}(\mathbf{k},\omega_{n}) \, .
\label{density-full}
\end{equation}
Here, the \emph{normal\/} single-particle Green's functions $\mathcal{G}_{11}^{\mathrm{pf}}$ (with the suffix pf standing for ``pairing fluctuations'') is taken of the form \cite{PPS-2004}:
\begin{eqnarray}
\mathcal{G}_{11}^{\mathrm{pf}}(\mathbf{k},\omega_{n}) & = & \frac{1}{ i \omega_{n} - \xi(\mathbf{k}) - \Sigma_{11}(\mathbf{k},\omega_{n}) - \frac{\Delta^{2}}{i \omega_{n} + \xi(\mathbf{k}) + \Sigma_{11}(\mathbf{k},-\omega_{n})}} 
\nonumber \\
\Sigma_{11}(\mathbf{k},\omega_{n}) \! & = & \! - \!\! \int \!\!\! \frac{d \mathbf{Q}}{(2 \pi)^{3}} \frac{1}{\beta} \sum_{\Omega_{\nu}} \Gamma_{11}(\mathbf{Q},\Omega_{\nu}) \, \mathcal{G}_{11}^{\mathrm{mf}}(\mathbf{Q}-\mathbf{k},\Omega_{\nu}-\omega_{n}) 
\nonumber \\
\mathcal{G}_{11}^{\mathrm{mf}}(\mathbf{k},\omega_{n}) & = & \frac{i \omega_{n} + \xi(\mathbf{k})}{(i \omega_{n} - E(\mathbf{k}))(i \omega_{n} + E(\mathbf{k}))}
\label{G_11-pairing-fluctuations}
\end{eqnarray}

\noindent
where $\Omega_{\nu} =  2\nu\pi/\beta$ ($\nu$ integer) is a bosonic Matsubara frequency \cite{FW-1971}, $\mathcal{G}_{11}^{\mathrm{mf}}$ the mean-field version of $\mathcal{G}_{11}$, and $\Gamma_{11}$ the $11$-component of the particle-particle ladder propagator to be discussed in detail below (cf. Eqs.~(\ref{full-particle-particle-ladder})-(\ref{B})).
The values of $\Delta$ and $\mu$ obtained in this way can then be entered into the mean-field expression (\ref{condensate-density-mean-field}) for $n_{0}$.
In this case, our results should be compared with those of Ref.~\cite{Griffin-2007} where the condensate density was obtained with the inclusion of pairing fluctuation effects within a Gaussian approximation, although only at first order in the relevant fluctuation contribution.

\noindent
(iii) Next, following the approach of Ref.~\cite{PPS-2018}, we will improve on the approach of Ref.~\cite{PPS-2004} and include the GMB correction throughout the BCS-BEC crossover at any temperature below $T_{c}$.
Accordingly, the density equation will be kept of the form (\ref{density-full}) while the gap equation will be cast in a form equivalent to a Hugenholtz-Pines condition for fermion pairs, to which the GMB correction can be added through (normal and anomalous) bosonic-like self-energies \cite{PPS-2018}.
The values of $\Delta$ and $\mu$ obtained in this way will again be entered into the mean-field expression (\ref{condensate-density-mean-field}) for $n_{0}$.

\noindent
(iv) Further improvements will be achieved by entering in the general expression (\ref{condensate-density-general}) for $n_{0}$ the \emph{anomalous\/} single-particle Green's function $\mathcal{G}_{21}^{\mathrm{pf}}$ taken from Ref.~\cite{PPS-2004}, namely,
\begin{eqnarray}
\mathcal{G}_{21}^{\mathrm{pf}}(\mathbf{k},\omega_{n}) = & \Delta & \left[ \left( i \omega_{n} - \xi(\mathbf{k}) - \Sigma_{11}(\mathbf{k},\omega_{n}) \right) \right.
\label{G_21-pairing-fluctuations} \\
& \times & \left. \left( i \omega_{n} + \xi(\mathbf{k}) + \Sigma_{11}(\mathbf{k},-\omega_{n}) \right) - \Delta^{2} \right]^{-1} 
\nonumber 
\end{eqnarray}

\noindent
whose form now includes pairing fluctuations beyond mean field.
Here, the values of $\Delta$ and $\mu$ can alternatively be taken at the mean-field level [point (i) above], with the inclusion of pairing fluctuations within the $t$-matrix [point (ii) above], and with the further inclusion of the GMB correction [point (iii) above].
In this way, we will be able to test how the replacement $\mathcal{G}_{21}^{\mathrm{mf}} \rightarrow \mathcal{G}_{21}^{\mathrm{pf}}$ in the expression (\ref{condensate-density-general}) will affect $n_{0}$, over and above the effects of taking the numerical values of $\Delta$ and $\mu$ at increasing levels of sophistication according to the points (i), (ii), and (iii) above.

The numerical results obtained in these ways will eventually be compared with those obtained by several Quantum Monte Carlo simulations as well as with the experimental data of Ref.~\cite{Roati-2020}.

\vspace{-0.5cm}
\subsection{Normal contribution to $h_{2}$}
\label{subsec:normal_contribution}
\vspace{-0.2cm}

With the definitions (\ref{grouping-spatial-coordinates}) and (\ref{grouping-spatial-coordinates-primed}) and the dictionary (\ref{compact-notation}), the second term on the right-hand side of Eq.~(\ref{Bethe-Salpeter-equation}) reads:
\begin{equation}
\mathcal{G}(1,1') \, \mathcal{G}(2,2') = \mathcal{G}_{11}(\mathbf{r}_{1'}-\mathbf{r}_{1},0^{-}) \, \mathcal{G}_{22}(\mathbf{r}_{2}-\mathbf{r}_{2'},0^{+}) 
\label{second-term-h_2-general}
\end{equation}
where again the ($\pm$) infinitesimals refer to the imaginary time domain \cite{footnote-1}.
In this term, one may set $\mathbf{r}_{1} = \mathbf{r}_{2}=\mathbf{r}$ and $\mathbf{r}_{1'} = \mathbf{r}_{2'}=\mathbf{r'}$ such that $\mathbf{r}_{1'}-\mathbf{r}_{1}=\mathbf{r}_{2'}-\mathbf{r}_{2}=\mathbf{R}$
(cf. \color{red}Fig.~\ref{Figure-1}(a)\color{black}), and study directly its dependence on $R = |\mathbf{R}|$.

In particular, at the mean-field level one obtains in Eq.~(\ref{second-term-h_2-general}):
\begin{eqnarray}
&& \mathcal{G}_{11}^{\mathrm{mf}}(\mathbf{R},0^{-}) \! = \!\! \int \!\!\! \frac{d \mathbf{k}}{(2 \pi)^{3}} e^{i \mathbf{k} \cdot \mathbf{R}} \frac{1}{\beta} \sum_{\omega_{n}} e^{i \omega_{n} \eta} \, \mathcal{G}_{11}^{\mathrm{mf}}(\mathbf{k},\omega_{n})
\label{G_11-mean-field} \\
& = & \int \!\!\! \frac{d \mathbf{k}}{(2 \pi)^{3}} e^{i \mathbf{k} \cdot \mathbf{R}} \left\{ u(\mathbf{k})^{2} f(E(\mathbf{k})) + v(\mathbf{k})^{2} \left[ 1 - f(E(\mathbf{k})) \right] \right\}
\nonumber
\end{eqnarray}

\noindent
where $u(\mathbf{k})^{2} = 1- v(\mathbf{k})^{2} = \frac{1}{2} \left( 1 + \frac{\xi(\mathbf{k})}{E(\mathbf{k})} \right)$ are the BCS coherence factors \cite{FW-1971}.
A similar result can be obtained for $\mathcal{G}_{22}^{\mathrm{mf}}(\mathbf{R},0^{+})$, by taking into account that, quite generally, $\mathcal{G}_{22}(\mathbf{k},\omega_{n}) = - \mathcal{G}_{11}(\mathbf{k},-\omega_{n})$.

For a non-interacting Fermi gas at zero temperature, on the other hand, the expression (\ref{G_11-mean-field}) would reduce to the form \cite{Mahan-1981} 
\begin{equation}
\mathcal{G}_{0}(R|T=0) = \frac{1}{2 \, \pi^{2}} \left[ \frac{\sin(k_{F} R)}{R^{3}} - k_{F} \frac{\cos(k_{F} R)}{R^{2}} \right] \, ,
\label{G_11-non-interacting} 
\end{equation}
which shows the characteristic Friedel's oscillations due to the sharpness of the Fermi surface.
At finite temperature, the Fermi surface is smeared by temperature effects and the amplitude of the oscillations decays exponentially \cite{FW-1971}.
For a gas with attractive inter-particle interaction, the Fermi surface is further smeared by interaction effects even at zero temperature.
In Appendix~\ref{sec:appendix-A}, it will be shown that $\mathcal{G}_{11}^{\mathrm{mf}}(\mathbf{R},0^{-})$ converges exponentially to zero for increasing $R$ over a length scale $\xi_{1}$;
in Sec.~\ref{subsec:results_normal}, it will further be shown that  $\xi_{1}$ (about) coincides with the Cooper pair size $\xi_{\mathrm{pair}}$ for any coupling throughout the BCS-BEC crossover, both at zero \cite{PS-1994} and finite \cite{PS-2014} temperature.
Similar results for the exponential damping of Friedel's oscillations were obtained in Ref.~\cite{Fetter-1965} both at $T=0$ and $T \rightarrow T_{c}^{-}$ but only in the BCS (weak-coupling) limit,
and more recently in Ref.~\cite{Romero-2020} throughout the BCS-BEC crossover but only at $T=0$.

\vspace{-0.5cm}
\subsection{Fluctuations contribution to $h_{2}$ \\ within the $t$-matrix approximation}
\label{subsec:fluctuations-contribution}
\vspace{-0.2cm}

The third term on the right-hand side of Eq.~(\ref{Bethe-Salpeter-equation}) represents the contribution $\delta h_{2}$ by pairing fluctuations to the two-particle reduced density matrix $h_{2}$.
This contribution, which survives even in the normal phase above $T_{c}$ where it reveals the presence of fluctuating Cooper pairs, is expected to progressively vanish for increasing distance $R$ between the centers of mass of the two pairs 
(cf. \color{red}Fig.~\ref{Figure-1}(a)\color{black}).
In the following, we shall analyze in detail the dependence of $\delta h_{2}$ on $R$ and extract from it, whenever possible, a characteristic (coupling and temperature dependent) length scale.
Or, at least, determine from it a definite distance at which the asymptotic correlations entailed by the ODLRO are effectively reached.

To this end, it will be convenient to eliminate at the outset the dependence of $\delta h_{2}$ on the relative coordinates $\boldsymbol{\rho}$ and $\boldsymbol{\rho}'$ of the separate pairs (cf. \color{red}Fig.~\ref{Figure-1}(a)\color{black})
and concentrate directly on the dependence of $\delta h_{2}$ on $R$.
This is achieved by introducing the so-called \emph{projected density matrix\/} \cite{Senatore-2002}
\begin{equation}
\delta h(\mathrm{R}) = \int \!\! d\boldsymbol{\rho} \, \delta h_{2}(\boldsymbol{\rho},\boldsymbol{\rho},\mathbf{R})
\label{projected-density-matrix}
\end{equation}
as far as the fluctuations contribution $\delta h_{2}$ is concerned.
[Note that a similar procedure was already followed in Eq.~(\ref{condensate-density-general}) to obtain the condensate density $n_{0}$ from the anomalous contribution to $h_{2}$.]

The simplest pairing fluctuations contribution $\delta h_{2}(\boldsymbol{\rho},\boldsymbol{\rho}',\mathbf{R})$ that one can consider is depicted diagrammatically in \color{red}Fig.~\ref{Figure-1}(c)\color{black}.
Taking into account the dictionary (\ref{compact-notation}) for the space and Nambu spin coordinates of the two-particle reduced density matrix, and owing to the contact form of the inter-particle interaction 
for the superfluid Fermi gas, the diagram of \color{red}Fig.~\ref{Figure-1}(c) \color{black} has the same topological structure of a Maki-Thompson (MT) diagram \cite{Thompson-1970,Maki-1970}.
Accordingly, it inherits the same overall sign of the MT diagram, due to the way the pairs of external points (1,1') and (2,2') therein are mutually connected.
The contribution to $\delta h_{2}(\boldsymbol{\rho},\boldsymbol{\rho}',\mathbf{R})$ from this diagram can then be cast in the following matrix form:
\begin{widetext}
\begin{equation}
\delta h_{2}(\boldsymbol{\rho},\boldsymbol{\rho'},\mathbf{R}) \! = \!\!\! \int \!\!\! \frac{d \mathbf{Q}}{(2 \pi)^{3}} \, e^{i \mathbf{Q} \cdot \left( \mathbf{R} + \frac{\boldsymbol{\rho}-\boldsymbol{\rho'}}{2} \right)} \frac{1}{\beta} \! \sum_{\nu} \! e^{i \Omega_{\nu} \eta} \!\!
\left[ \tilde{\Pi}_{11}(\boldsymbol{\rho};\mathbf{Q},\Omega_{\nu}),\tilde{\Pi}_{12}(\boldsymbol{\rho};\mathbf{Q},\Omega_{\nu}) \right] \!\! \left[ \!\! \begin{array}{cc} \Gamma_{11}(\mathbf{Q},\Omega_{\nu}) & \Gamma_{12}(\mathbf{Q},\Omega_{\nu}) \\ 
\Gamma_{21}(\mathbf{Q},\Omega_{\nu}) & \Gamma_{22}(\mathbf{Q},\Omega_{\nu}) \end{array} \! \right] \!\!\!
\left[ \!\! \begin{array}{c} \tilde{\Pi}_{11}(\boldsymbol{\rho'};\mathbf{Q},\Omega_{\nu}) \\ \tilde{\Pi}_{12}(\boldsymbol{\rho'};\mathbf{Q},\Omega_{\nu}) \end{array} \!\! \right] \, .
\label{explicit-form-delta-h_2} 
\end{equation}
Here,
\begin{eqnarray}
\tilde{\Pi}_{11}(\boldsymbol{\rho};\mathbf{Q},\Omega_{\nu}) & = & \! \int \!\!\! \frac{d \mathbf{k}}{(2 \pi)^{3}} \, e^{i \mathbf{k} \cdot \boldsymbol{\rho}} 
\frac{1}{\beta} \sum_{n} \mathcal{G}_{11}^{\mathrm{mf}}(\mathbf{k} + \mathbf{Q},\omega_{n}+\Omega_{\nu}) \, \mathcal{G}_{11}^{\mathrm{mf}}(\mathbf{k},-\omega_{n})
\label{Pi-tilde-11} \\
\tilde{\Pi}_{12}(\boldsymbol{\rho};\mathbf{Q},\Omega_{\nu}) & = & \! \int \!\!\! \frac{d \mathbf{k}}{(2 \pi)^{3}} \, e^{i \mathbf{k} \cdot \boldsymbol{\rho}} 
\frac{1}{\beta} \, \sum_{n} \mathcal{G}_{12}^{\mathrm{mf}}(\mathbf{k} + \mathbf{Q},\omega_{n}+\Omega_{\nu}) \, \mathcal{G}_{12}^{\mathrm{mf}}(\mathbf{k},-\omega_{n})
\label{Pi-tilde-12}
\end{eqnarray}
are form factors, and
\begin{equation}
\left[ \! \begin{array}{cc} \Gamma_{11}(\mathbf{Q},\Omega_{\nu}) & \Gamma_{12}(\mathbf{Q},\Omega_{\nu}) \\ \Gamma_{21}(\mathbf{Q},\Omega_{\nu}) & \Gamma_{22}(\mathbf{Q},\Omega_{\nu}) \end{array} \! \right]
= \frac{1}{ A(\mathbf{Q},\Omega_{\nu}) \, A(\mathbf{Q},-\Omega_{\nu}) - B(\mathbf{Q},\Omega_{\nu})^{2}}  
\left[ \! \begin{array}{cc} A(\mathbf{Q},-\Omega_{\nu}) & B(\mathbf{Q},\Omega_{\nu}) \\ B(\mathbf{Q},\Omega_{\nu}) & A(\mathbf{Q},\Omega_{\nu}) \end{array} \! \right] 
\label{full-particle-particle-ladder}
\end{equation}
are the components of the particle-particle ladder where
\begin{eqnarray}
A(\mathbf{Q},\Omega_{\nu}) \! & = & \! - \frac{m}{4 \pi a_{F}} + \int \! \frac{d \mathbf{k}}{(2 \pi)^{3}} \, \frac{m}{\mathbf{k}^{2}} 
- \int \!\!\! \frac{d \mathbf{k}}{(2 \pi)^{3}} \frac{1}{\beta} \! \sum_{n} \mathcal{G}_{11}^{\mathrm{mf}}(\mathbf{k} + \mathbf{Q},\omega_{n}+\Omega_{\nu}) \, \mathcal{G}_{11}^{\mathrm{mf}}(\mathbf{k},-\omega_{n})
\label{A} \\
B(\mathbf{Q},\Omega_{\nu}) \! & = & \!\! \int \!\!\! \frac{d \mathbf{k}}{(2 \pi)^{3}} \frac{1}{\beta} \! \sum_{n} \mathcal{G}_{12}^{\mathrm{mf}}(\mathbf{k} + \mathbf{Q},\omega_{n}+\Omega_{\nu}) \, \mathcal{G}_{12}^{\mathrm{mf}}(\mathbf{k},-\omega_{n})
\label{B}
\end{eqnarray}
\end{widetext}
according to the notation of Refs.~\cite{APS-2003,PPS-2004}.
In Eq.~(\ref{A}) use has been made of the regularization condition
\begin{equation}
\frac{m}{4 \pi a_{F}} = \frac{1}{v_{0}} + \int_{|\mathbf{k}| \le k_{0}} \! \frac{d \mathbf{k}}{(2 \pi)^{3}} \, \frac{m}{\mathbf{k}^{2}} \, ,
\label{regularization}
\end{equation}
whereby the limits $v_{0} \rightarrow 0^{-}$ for the strength of the contact inter-particle interaction and $k_{0} \rightarrow \infty$ for the ultraviolet cutoff are taken simultaneously, with $a_{F}$ kept at the desired value.
Note that in the normal phase above $T_{c}$, whereby $\mathcal{G}_{12}^{\mathrm{mf}}(\mathbf{k},\omega_{n}) = \Delta / \! \left( \omega_{n}^{2} + E(\mathbf{k})^{2} \right) \rightarrow 0$ and 
$\mathcal{G}_{11}^{\mathrm{mf}}(\mathbf{k},\omega_{n}) \rightarrow \mathcal{G}_{0}(\mathbf{k},\omega_{n})  = (\xi(\mathbf{k}) - i \omega_{n})^{-1}$, only the term containing 
$\Gamma_{11}(\mathbf{Q},\Omega_{\nu})  \rightarrow  A(\mathbf{Q},\Omega_{\nu})^{-1}$ survives in Eq.~(\ref{explicit-form-delta-h_2}) and the resulting expression corresponds to the non-self-consistent 
$t$-matrix approximation above $T_{c}$ \cite{Pini-2019}. 

The expression (\ref{explicit-form-delta-h_2}) gets considerably simplified by setting $\boldsymbol{\rho}=\boldsymbol{\rho'}$ and integrating over $\boldsymbol{\rho}$, to obtain the projected density matrix $\delta h(R)$ according to Eq.~(\ref{projected-density-matrix}).
The dependence on $R$ of the resulting expression for $\delta h(R)$ will be calculated numerically in Sec.~\ref{subsec:results_fluctuations}, for all temperatures in the superfluid phase and couplings
throughout the BCS-BEC crossover.

\vspace{-0.5cm}
\subsection{BEC limit of the fluctuations contribution to $h_{2}$}
\label{subsec:fluctuations-contribution-BEC}
\vspace{-0.2cm}

Before embarking in numerical calculations, it is relevant to show analytically that the fermionic expression for $\delta h(R)$, as it results from 
Eqs.~(\ref{projected-density-matrix}) and (\ref{explicit-form-delta-h_2}), reduces in the strong-coupling (BEC) limit to the fluctuations contribution to the one-particle density matrix for a gas of bosons described by the Bogoliubov approximation (further details will be given in Appendix~\ref{sec:appendix-B} at zero temperature).
This analysis appears important, not only because it represents a benchmark for the fully fermionic calculation of the projected density matrix when carried over to the BEC limit, but also because it highlights what we shall later find from our numerical results 
along the whole BCS-BEC crossover, about: 
(i) The progressive evolution for rising temperature of the long-range spatial behavior of the equal-time correlation function (\ref{two-particle-reduced-density-matrix}) into that of its static counterpart;
(ii) The identification of a single temperature-dependent length scale even in the presence of an asymptotic inverse-power-law spatial behavior.
[A corresponding analysis of the asymptotic spatial behavior of the fluctuations contribution to $\delta h(R)$  in the BCS (weak-coupling) limit will be considered in Appendix~\ref{sec:appendix-C} at zero temperature.]

In the BEC limit, whereby $\Delta/|\mu| << 1$ and $2m|\mu| \simeq a_{F}^{-2}$ with $\mu <0$ \cite{Physics-Reports-2018}, the following leading contribution to $\delta h(R)$ is obtained from
Eq.~(\ref{explicit-form-delta-h_2}):
\begin{eqnarray}
\delta h(R) & \simeq & \int \!\!\! \frac{d \mathbf{Q}}{(2 \pi)^{3}} \, e^{i \mathbf{Q} \cdot \mathbf{R}} \, \frac{1}{\beta} \sum_{\nu} e^{i \Omega_{\nu} \eta} \, \Gamma_{11}(\mathbf{Q},\Omega_{\nu}) 
\nonumber \\
& \times &  \int \!\! d\boldsymbol{\rho} \,\, \tilde{\Pi}_{11}(\boldsymbol{\rho};\mathbf{Q},\Omega_{\nu}) \, \tilde{\Pi}_{11}(-\boldsymbol{\rho};\mathbf{Q},\Omega_{\nu}) \, .
\label{dominant-term-BEC-limit}
\end{eqnarray}
In this expression, the form factor (\ref{Pi-tilde-11}) can be further approximated consistently with the BEC limit, yielding
\begin{eqnarray}
&& \int \!\! d\boldsymbol{\rho} \,\, \tilde{\Pi}_{11}(\boldsymbol{\rho};\mathbf{Q},\Omega_{\nu}) \, \tilde{\Pi}_{11}(-\boldsymbol{\rho};\mathbf{Q},\Omega_{\nu}) 
\label{form-factors-BEC-limit-1} \\
& = & \int \!\!\! \frac{d \mathbf{k}}{(2 \pi)^{3}} \left[ \frac{1}{\beta} \, \sum_{n} \mathcal{G}_{11}^{\mathrm{mf}}(\mathbf{k} + \mathbf{Q},\omega_{n}+\Omega_{\nu}) \, \mathcal{G}_{11}^{\mathrm{mf}}(\mathbf{k},-\omega_{n}) \right]^{2}
\nonumber \\
& \simeq & \!\! \int \!\!\! \frac{d \mathbf{k}}{(2 \pi)^{3}} \frac{1}{4 \, E(\mathbf{k})^{2}} \! \simeq \! \frac{m^{2}}{2 \, \pi^{2}} \!\! \int_{0}^{\infty} \!\!\!\!\! dk \frac{k^{2}}{\left( k^{2} + 2m|\mu| \right)^{2}} \simeq \! \frac{m^{2} \, a_{F}}{8 \, \pi} .
 \nonumber
\end{eqnarray}

\noindent
We also recall from Ref.~\cite{APS-2003} that, in the BEC limit, $\Gamma_{11}$ acquires the following form 
\begin{equation}
\Gamma_{11}(\mathbf{Q},\Omega_{\nu}) \simeq - \frac{8 \, \pi}{m^{2} \, a_{F}} \, \mathcal{G}^{'}_{B}(\mathbf{Q},\Omega_{\nu}) \, ,
\label{Gamma_11-BEC_limit}
\end{equation}
where
\begin{equation}
\mathcal{G}^{'}_{B}(\mathbf{Q},\Omega_{\nu}) = \frac{u_{B}(\mathbf{Q})^{2}}{i \Omega_{\nu} - E_{B}(\mathbf{Q})} - \frac{v_{B}(\mathbf{Q})^{2}}{i \Omega_{\nu} + E_{B}(\mathbf{Q})}
\label{bosonic-normal-propagator}
\end{equation}
is the normal bosonic single-particle propagator within the Bogoliubov approximation, with the bosonic coherence factors \cite{FW-1971}
\begin{equation}
u_{B}(\mathbf{Q})^{2} = 1+ v_{B}(\mathbf{Q})^{2} = \frac{1}{2} \, \left[ \frac{ \frac{\mathbf{Q}^{2}}{2 \, m_{B}} + \mu_{B} }{ E_{B}(\mathbf{Q}) } + 1 \right]
\label{Bogoliubov-u-and_v}
\end{equation}
and the dispersion relation
\begin{equation}
E_{B}(\mathbf{Q}) = \sqrt{ \left( \frac{\mathbf{Q}^{2}}{2 \, m_{B}} + \mu_{B} \right)^{2} - \mu_{B}^{2} } \,\, .
\label{Bogoliubov-E}
\end{equation}
In the above expressions, $m_{B} = 2m$ is the mass of the composite bosons (dimers) that form in the BEC limit and $\mu_{B} = \frac{4 \pi a_{B}}{m_{B}} n_{B}$ is the bosonic chemical potential obtained from
the definition $\mu_{B} = 2 \mu + (m a_{F}^{2})^{-1}$ in terms of the chemical potential $\mu$ of the constituent fermions \cite{Physics-Reports-2018}.
Here, $n_{B}=n/2$ and $a_{B} = 2 a_{F}$ are the bosonic density and scattering length expressed in terms of their fermionic counterparts $n$ and $a_{F}$, at the level of the present approximation.
With the results (\ref{form-factors-BEC-limit-1})-(\ref{bosonic-normal-propagator}), the expression (\ref{dominant-term-BEC-limit}) in the BEC limit becomes eventually:
\begin{eqnarray}
& & \delta h(R) \simeq - \int \!\!\! \frac{d \mathbf{Q}}{(2 \pi)^{3}} \, e^{i \mathbf{Q} \cdot \mathbf{R}} \, \frac{1}{\beta} \sum_{\nu} e^{i \Omega_{\nu} \eta} \, \mathcal{G}^{'}_{B}(\mathbf{Q},\Omega_{\nu})
\label{dominant-term-BEC-limit-final} \\
& = & \!\! \int \!\!\! \frac{d \mathbf{Q}}{(2 \pi)^{3}} e^{i \mathbf{Q} \cdot \mathbf{R}} \! \left\{ v_{B}(\mathbf{Q})^{2} + \left[ u_{B}(\mathbf{Q})^{2} \! + \! v_{B}(\mathbf{Q})^{2} \right] \! b(E_{B}(\mathbf{Q})) \! \right\}
\nonumber
\end{eqnarray}

\noindent
where $b(\epsilon) = \left( e^{\beta \epsilon} - 1 \right)^{-1}$ is the Bose function.

The results (\ref{bosonic-normal-propagator}) and (\ref{dominant-term-BEC-limit-final}) can now be exploited to illustrate in rather simple terms how the evolution with increasing temperature,
between the asymptotic spatial behaviors of the \emph{static\/} and \emph{equal-time\/} correlation functions, manifests itself in practice.

In the static limit, one sets $\Omega_{\nu}=0$ in the expression (\ref{bosonic-normal-propagator}) and obtains for all values of $\mathbf{Q}$
\begin{eqnarray}
- \, \mathcal{G}^{'}_{B}(\mathbf{Q},\Omega_{\nu}=0) & = & \frac{u_{B}(\mathbf{Q})^{2} + v_{B}(\mathbf{Q})^{2}}{E_{B}(\mathbf{Q})}
\nonumber \\
& = & m_{B} \! \left( \! \frac{1}{\mathbf{Q}^{2}} + \frac{1}{\mathbf{Q}^{2} + 4 m_{B} \mu_{B} } \! \right) ,
\label{bosonic-normal-propagator-static}
\end{eqnarray}
where $4 m_{B} \mu_{B} = \xi_{B}^{-2}$ defines the \emph{healing length} $\xi_{B}$ in the BEC limit.
The two terms within parentheses on the right-hand side of Eq.~(\ref{bosonic-normal-propagator-static}) correspond, respectively, to the transverse (massless) and longitudinal (massive) contributions to the static correlation function.
Both contributions were considered in Ref.~\cite{PS-1996} at zero temperature, and the longitudinal contribution was considered in Ref.~\cite{PS-2014} also at finite temperature (up and past $T_{c}$),
with the purpose of identifying the healing length at any temperature for a fermionic system undergoing the BCS-BEC crossover.
In the BEC limit, the two terms on the right-hand side of Eq.~(\ref{bosonic-normal-propagator-static}) then yield the following behavior in real space
\begin{equation}
- \mathcal{G}^{'}_{B}(\mathbf{R},\Omega_{\nu}=0) = \frac{m_{B}}{4 \pi R} \left( 1 + e^{-R/\xi_{B}} \right)
\label{bosonic-normal-propagator-static-asymptotic}
\end{equation}
with $R = |\mathbf{R}|$.
This result shows how $\xi_{B}$ can be identified from the exponentially decaying term, which is associated with the longitudinal contribution. 

For the equal-time function, on the other hand, one obtains from the expression (\ref{bosonic-normal-propagator}) in the zero-temperature limit
\begin{eqnarray}
& - & \frac{1}{\beta} \sum_{\nu} e^{i \Omega_{\nu} \eta} \, \mathcal{G}^{'}_{B}(\mathbf{Q},\Omega_{\nu}) \,\, \stackrel{(T \rightarrow 0)}{-\!-\!\!\! \longrightarrow} \,\, v_{B}(\mathbf{Q})^{2}
\nonumber \\
& = & \frac{1}{4} \! \left( \frac{ \sqrt{1 + Q^{2} \xi_{B}^{2}} }{ Q \xi_{B}} - 1 \right) + \frac{1}{4} \! \left( \frac{ Q \xi_{B}}{ \sqrt{1 + Q^{2} \xi_{B}^{2}} } - 1 \right)
\nonumber \\
& & \stackrel{(Q \xi_{B} \ll 1)}{-\!-\!\!\! \longrightarrow} - \frac{1}{2} + \frac{1}{4 \xi_{B} Q} + \frac{3}{8} \, Q \xi_{B} + \cdots
\label{bosonic-normal-propagator-equal-time-low-T}
\end{eqnarray}
with $Q = |\mathbf{Q}|$, where we have considered only the small-$Q$ behavior to capture directly the large-$R$ behavior of the corresponding Fourier transform in real space \cite{Lighthill-1959}.
Accordingly,
\begin{equation}
\!\!\!\! - \frac{1}{\beta} \sum_{\nu} e^{i \Omega_{\nu} \eta} \mathcal{G}^{'}_{B}(\mathbf{R},\Omega_{\nu}) \approx \frac{1}{8 \pi^{2} \xi_{B} R^{2}} \, + \, \mathcal{O} \! \left( \frac{1}{R^{4}} \right) \, + \, \cdots 
\label{bosonic-normal-propagator-equal-time-asymptotic}
\end{equation}
where $\xi_{B}$ is formally the \emph{same} length entering Eq.~(\ref{bosonic-normal-propagator-static-asymptotic}).

When raising the temperature close enough to $T_{c}$ such that $\mu_{B} \ll k_{B} T$, and for $Q$ small enough such that $Q^{2}/(2 m_{B}) \ll k_{B} T$, 
the Bose function $b(E_{B}(\mathbf{Q}))$ in Eq.~(\ref{dominant-term-BEC-limit-final}) becomes approximately $k_{B} T/ E_{B}(\mathbf{Q})$.
With the help of Eq.~(\ref{bosonic-normal-propagator-static}), one then obtains:
\begin{eqnarray}
& - & \! \frac{1}{\beta} \! \sum_{\nu} \! e^{i \Omega_{\nu} \eta} \mathcal{G}^{'}_{B}(\mathbf{Q},\Omega_{\nu}) 
\longrightarrow k_{B} T \! \left( \!\! \frac{u_{B}(\mathbf{Q})^{2} + v_{B}(\mathbf{Q})^{2}}{E_{B}(\mathbf{Q})} \!\! \right)
\label{bosonic-normal-propagator-equal-time-high-T} \\
& = & m_{B} k_{B} T \! \left( \!\! \frac{1}{\mathbf{Q}^{2}} + \frac{1}{\mathbf{Q}^{2} + 4 m_{B} \mu_{B} } \!\! \right) \! = \! - k_{B} T \mathcal{G}^{'}_{B}(\mathbf{Q},\Omega_{\nu}=0) .
\nonumber
\end{eqnarray}

\noindent
Note that the result on the right-hand side of this expression is equivalent to having retained only the terms with $\Omega_{\nu}=0$ in the original sum on the left-hand side of the same expression.
Accordingly, the asymptotic spatial behavior of $- \frac{1}{\beta} \sum_{\nu} e^{i \Omega_{\nu} \eta} \mathcal{G}^{'}_{B}(\mathbf{R},\Omega_{\nu})$ for large $R$ evolves into that given by Eq.~(\ref{bosonic-normal-propagator-static-asymptotic}), in agreement with what we had anticipated in the Introduction on general grounds.

The above evolution for increasing temperature, of the long-range spatial behavior of the equal-time correlation function into that of its static counterpart, will be recovered throughout the BCS-BEC crossover 
by the numerical calculations presented in the next Section for the projected density matrix given by Eqs.~(\ref{projected-density-matrix}) and (\ref{explicit-form-delta-h_2}).

In this context, it is convenient to summarize how the asymptotic spatial behavior of the projected density matrix (\ref{dominant-term-BEC-limit-final}) in the BEC limit evolves continuously from zero up to the critical temperature.
To this end, it is sufficient to determine the evolution with temperature of the small-$Q$ behavior of the Fourier transform of the expression (\ref{dominant-term-BEC-limit-final}), namely,
\begin{equation}
\delta h(\mathbf{Q}) = v_{B}(\mathbf{Q})^{2} + \left[ u_{B}(\mathbf{Q})^{2} \! + \! v_{B}(\mathbf{Q})^{2} \right] \! b(E_{B}(\mathbf{Q})) \, .
\label{dominant-term-BEC-limit-Q-space} 
\end{equation}
Three main regimes can be identified to the purpose:

\noindent
1) A low-temperature regime, whereby $k_{B} T \ll \frac{ \mathbf{Q}^{2} }{2m_{B}} \ll \frac{1}{m_{B} \xi_{B}^{2}}$.
In this case, 
\begin{equation}
\delta h(\mathbf{Q}) \simeq v_{B}(\mathbf{Q})^{2} \simeq \frac{1}{4 \xi_{B} Q}
\label{first-term-approximate-low}
\end{equation}
like in Eq.~(\ref{bosonic-normal-propagator-equal-time-low-T}), yielding $\delta h(\mathbf{R}) \simeq \frac{1}{8 \pi^{2} \xi_{B} R^{2}}$ like in Eq.~(\ref{bosonic-normal-propagator-equal-time-asymptotic}).
Accordingly, the value of $\xi_{B}$ can be obtained from this expression in the low-temperature regime.

\noindent
2) An intermediate-temperature regime, whereby $\frac{ \mathbf{Q}^{2} }{2m_{B}} \ll  k_{B} T \lesssim \frac{1}{m_{B} \xi_{B}^{2}}$, such that $E_{B}(\mathbf{Q}) \simeq \frac{|\mathbf{Q}|}{2 m_{B} \xi_{B}} \ll k_{B} T$.
In this case, the Bose function can be approximated by
\begin{equation}
b(E_{B}(\mathbf{Q})) \simeq \frac{k_{B}T}{E_{B}(\mathbf{Q})} - \frac{1}{2} + \cdots 
\label{approximate-Bose-function}
\end{equation}
such that the second term on the right-hand side of Eq.~(\ref{dominant-term-BEC-limit-Q-space}) becomes
\begin{equation}
\left[ u_{B}(\mathbf{Q})^{2} \! + \! v_{B}(\mathbf{Q})^{2} \right] \! b(E_{B}(\mathbf{Q})) \simeq \frac{m_{B} k_{B} T}{Q^{2}} - \frac{1}{4 \xi_{B} Q} \, .
\label{second-term-approximate}
\end{equation}
Here, the term $\propto Q^{-1}$ cancels the term of Eq.~(\ref{first-term-approximate-low}), thereby converting the leading asymptotic behavior of $\delta h(\mathbf{R})$ from $\mathcal{O}(R^{-2})$ to $\mathcal{O}(R^{-1})$.
Correspondingly, the relevant length turns out to be the \emph{thermal length\/} $\xi_{\mathrm{T}} = (m_{B} k_{B} T)^{-1/2}$, with no reference, however, to the intrinsic parameters of the many-body system.
In this temperature regime there is thus no way to extract the healing length $\xi_{B}$ from the spatial profile of $\delta h(\mathbf{R})$.

\noindent
3) A high-temperature regime, whereby $\frac{ \mathbf{Q}^{2} }{2m_{B}} \ll k_{B} T$ and $\frac{1}{m_{B} \xi_{B}^{2}} \ll k_{B} T$.
In this case, $E_{B}(\mathbf{Q}) \ll k_{B} T$ and $b(E_{B}(\mathbf{Q})) \simeq k_{B} T / E_{B}(\mathbf{Q})$ with $E_{B}(\mathbf{Q}) $ given by Eq.~(\ref{Bogoliubov-E}).
This yields approximately
\begin{eqnarray}
\delta h(\mathbf{Q}) & \simeq & k_{B} T \, \left[ \frac{u_{B}(\mathbf{Q})^{2} \! + \! v_{B}(\mathbf{Q})^{2}}{E_{B}(\mathbf{Q})} \right]
\nonumber \\
& = & m_{B} k_{B} T \left( \! \frac{1}{Q^{2}} + \frac{1}{Q^{2} + \xi_{B}^{-2} } \! \right) 
\label{first-term-approximate-high}
\end{eqnarray}
where the result (\ref{bosonic-normal-propagator-static}) has been utilized.
Correspondingly, the leading asymptotic behavior of $\delta h(\mathbf{R})$ becomes that of Eq.~(\ref{bosonic-normal-propagator-static-asymptotic}) (apart from an overall factor $k_{B} T $).

\begin{figure}[t]
\begin{center}
\includegraphics[width=8.0cm,angle=0]{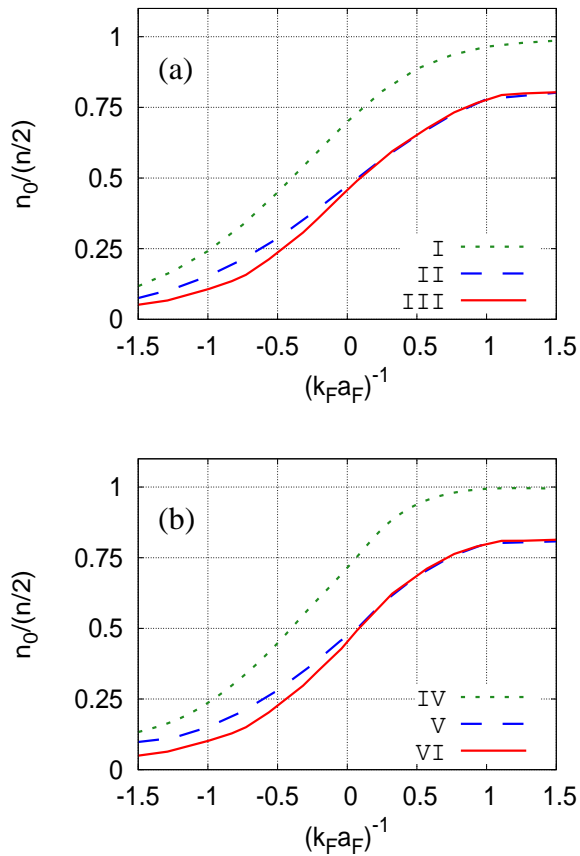}
\caption{(Color online) The condensate fraction $n_{0}$ (in units of half the number density $n$), calculated at zero temperature 
              within the six alternative approximations I-VI described in the text, is shown as a function of coupling. 
              In panel (a) the anomalous single-particle Green's function $\mathcal{G}_{21}$ is taken at the mean-field level, while in panel (b) $\mathcal{G}_{21}$
              also includes pairing fluctuations beyond mean field. 
              In addition, in both panels the values of the thermodynamic parameters $(\Delta,\mu)$ correspond to three different approximations (see the text).}
\label{Figure-2}
\end{center} 
\end{figure} 


\vspace{-0.4cm}
\section{Numerical results} 
\label{sec:numerical_results}
\vspace{-0.2cm}

In this Section, we present the numerical results obtained by evaluating all three terms on the right-hand side of Eq.~(\ref{Bethe-Salpeter-equation}) within the approximations that were specified in Secs.~\ref{subsec:anomalous_contribution}, \ref{subsec:normal_contribution}, and \ref{subsec:fluctuations-contribution}, respectively, over a wide range of coupling across the BCS-BEC crossover and temperature in the superfluid phase.
In the next three subsections, we shall deal with these three terms separately.

\vspace{-0.4cm}
\subsection{Results for the anomalous contribution}
\label{subsec:results_anomalous}
\vspace{-0.2cm}

\begin{figure}[t]
\begin{center}
\includegraphics[width=9.0cm,angle=0]{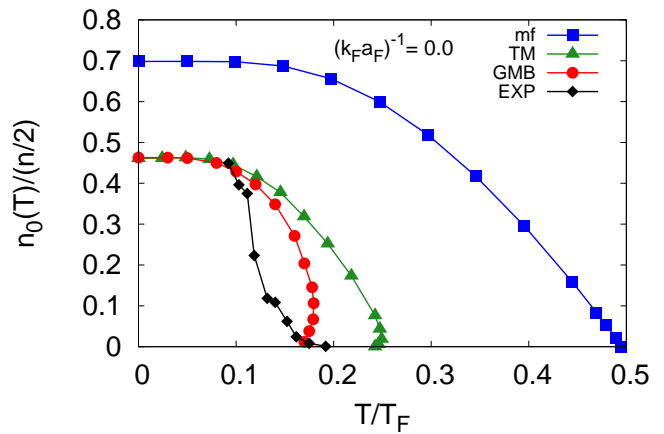}
\caption{(Color online) Condensate fraction $n_{0}(T)$ (in units of half the number density $n$) vs temperature $T$ (in units of the Fermi temperature $T_{F}$), obtained at unitarity 
                                     within the following three approximations described in the text: 
                                     I (mf - squares), V (TM - triangles), and VI (GMB - dots).
                                     The experimental results from Ref.~\cite{Zwierlein-2012} for the unitary Fermi gas are also shown for comparison (EXP - diamonds).}
\label{Figure-3}
\end{center} 
\end{figure} 

The relevant quantity to be extracted from the anomalous contribution to the two-particle reduced density matrix $h_{2}$ is the condensate density $n_{0}$ given by the expression (\ref{condensate-density-general}), which depends on the specific choice of the anomalous single-particle Green's function $\mathcal{G}_{21}$.

In this respect, several choices of $\mathcal{G}_{21}$ are at our disposal, as already discussed in detail in Sec.~\ref{subsec:anomalous_contribution}.
Let's here recall that we can either take $\mathcal{G}_{21}$ at the mean-field level (cf. Eq.~(\ref{G_12-mean-field})), with the thermodynamic parameters $(\Delta,\mu)$ calculated alternatively at the mean-field level (approximation I), within the $t$-matrix approach (approximation II),
and with the further inclusion of the GMB correction (approximation III).
Or else, we can take $\mathcal{G}_{21}$ of the form (\ref{G_21-pairing-fluctuations}) which includes pairing fluctuations beyond mean field, again with the thermodynamic parameters $(\Delta,\mu)$ calculated alternatively at the mean-field level 
(approximation IV), within the $t$-matrix approach (approximation V), and with the further inclusion of the GMB correction (approximation VI).
These calculations can be done for any coupling throughout the BCS-BEC crossover and for any temperature in the superfluid phase.

\begin{figure}[t]
\begin{center}
\includegraphics[width=9.0cm,angle=0]{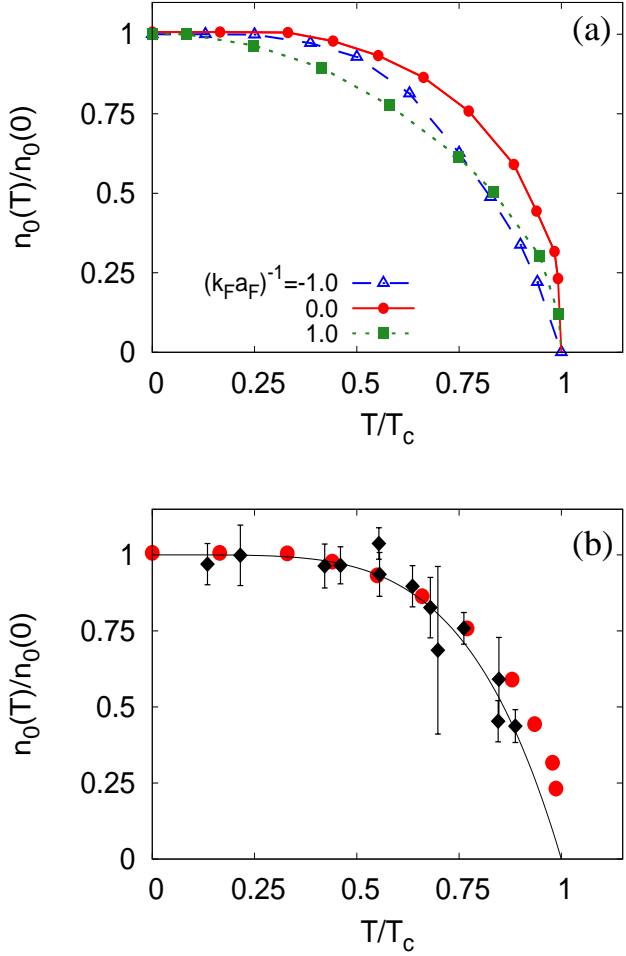}
\caption{(Color online) Condensate fraction $n_{0}(T)$ (in units of the value $n_{0}(0)$ at zero temperature) vs temperature $T$ (in units of the critical temperature $T_{c}$). 
                                    In panel (a), the results obtained within the (GMB) approximation VI are shown for three different couplings. 
                                    In panel (b), the results of the GMB calculation at unitarity (dots) are compared with the experimental data for $^{4}$He from Ref.~\cite{Prisk2017} (diamonds).
                                    The latter data are also fitted by the expression $1-(T/T_{c})^{\alpha}$ with $\alpha = 4.56$ (full line).}
\label{Figure-4}
\end{center} 
\end{figure} 

\color{red}Figure~\ref{Figure-2} \color{black} shows the results for the condensate density $n_{0}$ at zero temperature as a function of coupling spanning the unitary regime from the BCS to the BEC limits, obtained within the above six approximations I-VI.
Note how the main differences in these plots among the different curves are due to the use of the alternative sets of values for the thermodynamic parameters $(\Delta,\mu)$,
and not to the different functional forms used for the anomalous single-particle Green's function $\mathcal{G}_{21}$.
In the following, we shall consider approximation VI as the most sophisticated of our approximations, to the extent that it includes the GMB correction both in the functional form of $\mathcal{G}_{21}$ and in the parameters $(\Delta,\mu)$, and use it when comparing with the available experimental and Quantum Monte Carlo data.

With these premises, \color{red}Fig.~\ref{Figure-3} \color{black} shows (on an absolute temperature scale set by the Fermi temperature $T_{F}$) the temperature dependence of the condensate density $n_{0}(T)$ at unitarity, obtained within the mean-field approximation I (mf), 
the $t$-matrix approximation V (TM), and the approximation VI that includes also the GMB correction (GMB), as described above.
The experimental results from Ref.~\cite{Zwierlein-2012} for the unitary Fermi gas are also reported for comparison.
Close to the critical temperature, note the concave behavior of all theoretical curves, as well as the occurrence of a slight re-entrant behavior of  $n_{0}(T)$ within both the TM and GMB approaches, which is inherited from a similar behavior present in the gap parameter $\Delta(T)$ \cite{PPS-2018}.
On the other hand, the experimental results show an opposite (convex) behavior, which is rather peculiar for the temperature dependence of the condensate density and might originate from the rapid ramp method utilized in the experiment \cite{Zwierlein-2012}.

\begin{figure}[t]
\begin{center}
\includegraphics[width=6.3cm,angle=-90]{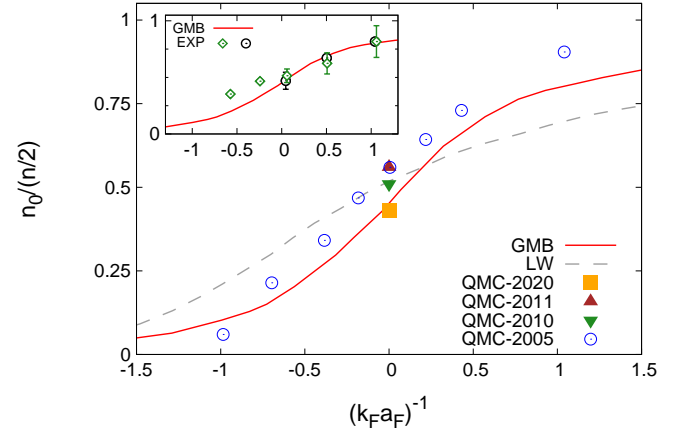}
\caption{(Color online) The coupling dependence of the condensate fraction $n_{0}$ at zero temperature (in units of half the number density $n$) obtained by our (GMB) approximation VI (full line) is compared with the Luttinger-Ward (LW) results of Ref.~\cite{Haussmann-2007}
                                    (dashed line) and with the Quantum Monte Carlo calculations of Refs.~\cite{Astrakharchik-2005} (QMC-2005, circles), \cite{Needs-2010} (QMC-2010, down triangles), \cite{Mitas-2011} 
                                    (QMC-2011, up triangles), and \cite{Lee-2020} (QMC-2020, squares).
                                    In the inset, the GMB results (full line) are further compared with the experimental data from Ref.~\cite{Roati-2020} (EXP, diamonds and circles).}
\label{Figure-5}
\end{center} 
\end{figure} 

\color{red}Figure~\ref{Figure-4}(a) \color{black} further shows the temperature dependence of the condensate density $n_{0}(T)$ for three different couplings, as obtained by our most sophisticated approximation VI (GMB) described above.
In addition, \color{red}Fig.~\ref{Figure-4}(b) \color{black} reproduces from panel (a) the temperature dependence of $n_{0}(T)$ at unitarity, and compares it with the temperature dependence of the condensate density obtained experimentally for $^{4}$He in Ref.~\cite{Prisk2017}.
In both panels, the condensate density is normalized to its value $n_{0}(0)$ at zero temperature, while the temperature is rescaled in terms of the respective value of the critical temperature $T_{c}$ for given coupling \cite{footnote-2}.
\color{red}Figure~\ref{Figure-4}(b) \color{black} shows also a least-square fit of the form $n_{0}(T)/n_{0}(0) = (1-(T/T_{c})^{\alpha})$ (full line) to the experimental data of Ref.~\cite{Prisk2017} (diamonds), for which we find $\alpha = 4.56$ \cite{Sears-1982}.
Note how this fit encompasses reasonably well also the values calculated at unitarity within the GMB approximation (dots).
Note also that to the temperature dependence of the experimental data for $^{4}$He reported in \color{red}Fig.~\ref{Figure-4}(b) \color{black} there corresponds a concave-type behavior, in contrast to the convex-type behavior of the experimental data from Ref.~\cite{Zwierlein-2012} for the unitary Fermi gas reported in \color{red}Fig.~\ref{Figure-3}\color{black}. 

Finally, \color{red}Fig.~\ref{Figure-5} \color{black} compares the results of our most sophisticated calculation for the condensate density $n_{0}$ (corresponding to the (GMB) approximation VI) across the unitary regime at zero temperature, with the results of a self-consistent $t$-matrix approach (LW) and of several Quantum Monte  Carlo (QMC) calculations.
Here, the smaller (larger) values of $n_{0}$ on the BCS (BEC) side of the crossover, obtained by the GMB approach with respect to 
those obtained by the LW approach, are consistent with a similar behavior obtained by the two approaches for the coupling dependence 
of the gap parameter $\Delta$ at zero temperature, as shown in Fig.~4 of Ref.~\cite{Moritz-2021}.
In that figure it is also apparent that on the BCS side of the crossover the experimental data for $\Delta$ agree better with the GMB than
with the LW results, while the opposite is true on the BEC side of the crossover.
The inset of \color{red}Fig.~\ref{Figure-5} \color{black} compares further the GMB results with the recent experimental data for $n_{0}$ 
from Ref.~\cite{Roati-2020}, for which a disagreement can be noted on the BCS side of the crossover.
In the light of the good agreement obtained on the BCS side of the crossover between the results of the GMB calculation 
with the experimental data for $\Delta$ reported in Ref.~\cite{Moritz-2021}, the disagreement for $n_{0}$ that appears in the inset of \color{red}Fig.~\ref{Figure-5} \color{black} on the BCS side of the crossover may possibly be due to the specific protocol adopted in Ref.~\cite{Roati-2020} to extract the values of $n_{0}$, which is expected to better apply to the BEC rather than to the BCS side of the crossover.

\begin{figure}[t]
\begin{center}
\includegraphics[width=7.0cm,angle=0]{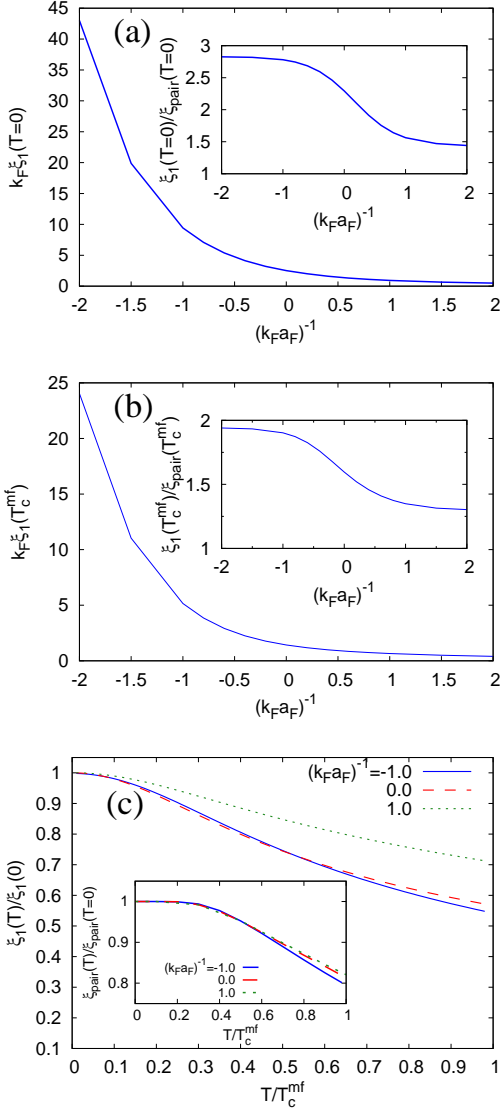}
\caption{(Color online) The coupling dependence of the length scale $\xi_{1}$ (in units of the inverse Fermi wave vector $k_{F}$), entering the normal contribution to $h_{2}$ and calculated at the mean-field level, is shown (a) at zero temperature and (b) at the critical temperature. 
                                    In both cases, the insets show a comparison with the corresponding coupling dependence of the Cooper pair size $\xi_{\mathrm{pair}}$ taken from Ref.~\cite{PS-2014}. 
                                    (c) The temperature dependence of $\xi_{1}$ (in units of the value $\xi_{1}(0)$ at zero temperature and of the mean-field critical temperature $T_{c}^{\mathrm{mf}}$) is shown for three characteristic couplings across unitarity. 
                                    The inset shows the corresponding temperature dependence of $\xi_{\mathrm{pair}}$ reproduced from Fig.~2(b) of Ref.~\cite{PS-2014}.}
\label{Figure-6}
\end{center} 
\end{figure} 

\vspace{-0.4cm}
\subsection{Results for the normal contribution}
\label{subsec:results_normal}
\vspace{-0.2cm}

The normal contribution to $h_{2}$, as given at the mean-field level by the expression (\ref{G_11-mean-field}), is considered in detail in Appendix~\ref{sec:appendix-A}, where its spatial behavior is conveniently expressed by the sum of elementary functions (cf. Eq.(\ref{resulting-expression-final}) therein).
In addition, in Appendix~\ref{sec:appendix-A} the overall spatial behavior of the expression (\ref{G_11-mean-field}) is shown to be suitably captured by a characteristic length $\xi_{1}$ given by Eq.~(\ref{xi_1}), which itself depends on coupling and temperature.
Here, we consider explicitly this dependence and compare it with that of the characteristic length $\xi_{\mathrm{pair}}$ entering the anomalous contribution to $h_{2}$, given at the mean-field level 
by the expression (\ref{G_12-mean-field}).
We recall that $\xi_{\mathrm{pair}}$ represents the \emph{intra-pair} correlation length (or Cooper pair size) and was calculated across the BCS-BEC crossover, at zero temperature in Ref.~\cite{PS-1994} and as a function of temperature in Ref.~\cite{PS-2014}. 

\color{red}Figure~\ref{Figure-6}(a) \color{black} shows the coupling dependence of $\xi_{1}$ at zero temperature, which is compared with the corresponding dependence of $\xi_{\mathrm{pair}}$ in the inset.
This comparison confirms that $\xi_{1}$ differs from $\xi_{\mathrm{pair}}$ by at most a factor of order unity, as it could have been expected on physical grounds.
In particular, at zero temperature one readily obtains from the expression (\ref{xi_1}) that $\xi_{1} \simeq k_{F}/(m \Delta) \simeq 2 \sqrt{2} \, \xi_{\mathrm{pair}}$ in the BCS limit 
(whereby $\Delta \ll \mu$ and $\mu \simeq E_{F}$), and that $\xi_{1} \simeq a_{F} \simeq \sqrt{2} \, \xi_{\mathrm{pair}}$ in the BEC limit (whereby $\Delta \ll |\mu|$ and $\mu \simeq - (2 m a_{F}^{2})^{-1}$).
As shown in the inset of \color{red}Fig.~\ref{Figure-6}(a)\color{black}, both these limiting values are reached essentially at the boundaries of the unitarity regime $-1 \lesssim (k_{F}\, a_{F})^{-1} \lesssim +1$ \cite{Ketterle-2008}.
\color{red}Figure~\ref{Figure-6}(b) \color{black} shows further the coupling dependence of $\xi_{1}$ at the (mean-field) critical temperature $T_{c}^{\mathrm{mf}}$,
which is again compared with the corresponding dependence of $\xi_{\mathrm{pair}}$ in the inset.
Even in this case, $\xi_{1}$ differs from $\xi_{\mathrm{pair}}$ by at most a factor of order unity.
Finally, \color{red}Fig.~\ref{Figure-6}(c) \color{black} shows the temperature dependence of $\xi_{1}$ for three characteristic couplings across the BCS-BEC crossover, while the inset reports for comparison the corresponding dependence of $\xi_{\mathrm{pair}}$ 
reproduced from Fig.~2(b) of Ref.~\cite{PS-2014}.

Due to the above similarities between the coupling and temperature dependences of $\xi_{1}$ and $\xi_{\mathrm{pair}}$, we expect that the inclusion of pairing fluctuations beyond mean field would only 
marginally affect $\xi_{1}$, similarly to what was explicitly shown to occur for $\xi_{\mathrm{pair}}$ in Ref.~\cite{PS-2014}.

\vspace{-0.4cm}
\subsection{Results for the fluctuations contribution}
\label{subsec:results_fluctuations}
\vspace{-0.2cm}

To obtain the leading asymptotic spatial behavior of the projected density matrix $\delta h(\mathbf{R})$ given by Eqs.~(\ref{projected-density-matrix}) and (\ref{explicit-form-delta-h_2}) throughout the BCS-BEC crossover, we follow closely the short summary made at the end of Sec.~\ref{subsec:fluctuations-contribution-BEC} for the corresponding behavior in the BEC limit and adapt it to the present context.
In this respect, we are implicitly assuming that, as far as the leading asymptotic spatial behavior of $\delta h(\mathbf{R})$ is concerned (and apart from the explicit numerical values of the healing length 
$\xi_{\mathrm{odlro}}$ associated with the ODLRO), there should be no substantial difference in the asymptotic functional dependence of $\delta h(\mathbf{R})$ on $R$ from the BEC to the BCS limits. 
This is \emph{provided\/} the considered values of $R$ are sufficiently larger than the spatial extent of the fermionic pairs involved in superfluidity, being either Cooper pairs in the BCS limit or composite bosons in the BEC limit.

Accordingly, as we did in Sec.~\ref{subsec:fluctuations-contribution-BEC}, we identify several temperature regimes where different asymptotic functional dependences of $\delta h(\mathbf{R})$ on $R$ are assumed to hold:

\noindent
1) A low-temperature regime $k_{B} T \ll \frac{1}{m \, \xi_{\mathrm{odlro}}^{2}}$, where we take
\begin{equation}
\delta h(\mathbf{R}) \approx \frac{1}{8 \pi^{2} \, \xi_{\mathrm{odlro}} \, R^{2}} \, ,
\label{asymptotic-regime-1}
\end{equation}
from which $\xi_{\mathrm{odlro}}$ can be readily extracted.
With this definition, the value of $\xi_{\mathrm{odlro}}$ coincides with the corresponding value of $\xi_{\mathrm{phase}}$ in the BEC regime.

\noindent
2) An intermediate-temperature regime $k_{B} T \lesssim \frac{1}{m \, \xi_{\mathrm{odlro}}^{2}}$ where we take
\begin{equation}
\delta h(\mathbf{R}) \approx \left\{ \begin{array}{cc}        \frac{1}{8 \pi^{2} \, \xi_{\mathrm{odlro}} \, R^{2}}    &   \,\,\, (R \lesssim \tilde{R})      \\
                                                                                                                                                                                                                           \\
                                                                                         c_{1} \, \frac{m k_{B} T}{2 \pi R}                             &   \,\,\, (R \gtrsim \tilde{R})       \end{array}  \right. 
\label{asymptotic-regime-2}
\end{equation}
where $c_{1}$ is a numerical coefficient (of order unity) and $\tilde{R}$ is consistently determined by the condition $c_{1} k_{B} T = (4 \pi \, m \, \xi_{\mathrm{odlro}} \, \tilde{R})^{-1}$.
Note, however, that the $R^{-2}$ behavior in Eq.~(\ref{asymptotic-regime-2}) may hardly be visible in practice,
to the extent that the (upper) limit of this intermediate-temperature regime corresponds to the condition $\xi_{\mathrm{odlro}} / \tilde{R} \lesssim 4 \pi$.
This implies that, in the temperature regime ``intermediate'' between $T \ll T_{c}$ and $T \lesssim T_{c}$, it is not possible to extract from the radial profile of $\delta h(\mathbf{R})$ a characteristic length 
$\xi_{\mathrm{odlro}}$ that depends on the parameters of the many-body system, a conclusion which is in line with that reached in Ref.~\cite{AT-1966} under a related perspective.

\noindent
3) A high-temperature regime $\frac{1}{m \, \xi_{\mathrm{odlro}}^{2}} \ll k_{B} T$ still in the superfluid phase, where we take
\begin{equation}
\delta h(\mathbf{R}) \approx c_{1} \, \frac{m k_{B} T}{2 \pi R} \left( 1 + c_{2} \, e^{-R/\xi_{\mathrm{odlro}}} \right) \, .
\label{asymptotic-regime-3}
\end{equation}
Here, the coefficient $c_{1}$ is determined for sufficiently large $R$ when the decaying exponential becomes negligible, while the coefficient $c_{2}$ and $\xi_{\mathrm{odlro}}$ are determined at smaller $R$ from this decaying exponential. 

\noindent
4) An even higher temperature regime $\frac{1}{m \, \xi_{\mathrm{odlro}}^{2}} \ll k_{B} T$ past $T_{c}$ in the normal phase, where we take
\begin{equation}
\delta h(\mathbf{R}) \approx c_{1} \, \frac{m k_{B} T}{ \pi R} \, e^{-R/\xi_{\mathrm{odlro}}}  \, ,
\label{asymptotic-regime-4}
\end{equation}
in line with a general argument on the asymptotic behavior of correlation functions in the neighbourhood of $T_{c}$ \cite{LeBellac-1995}.

The dimensionless coefficients $c_{1}$ and $c_{2}$ depend on coupling and on temperature within the respective temperature ranges.
In particular, from the analysis of Sec.~\ref{subsec:fluctuations-contribution-BEC} both coefficients are expected to tend to unity in the BEC limit.
In the following, we shall omit reporting the values of the coefficients $c_{1}$ and $c_{2}$ and rather concentrate on the length scale $\xi_{\mathrm{odlro}}$ of direct physical interest.

\begin{figure}[t]
\begin{center}
\includegraphics[width=9.0cm,angle=0]{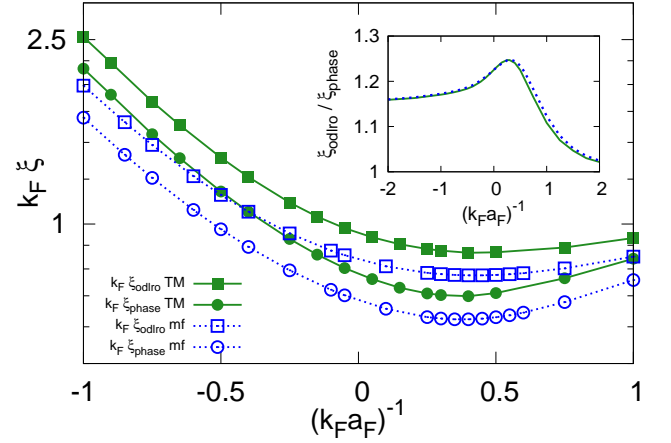}
\caption{(Color online) The coupling dependence of $\xi_{\mathrm{odlro}}$ at zero temperature (squares) is compared with that of $\xi_{\mathrm{phase}}$ (circles). 
                                    These two lengths (given in units of the inverse Fermi wave vector $k_{F}$) are computed with the thermodynamic parameters, calculated either at the mean-field level 
                                    (empty symbols and dotted lines) or with the inclusion of pairing fluctuations at the level of the $t$-matrix approach (filled symbols and full lines).
                                    In the inset, the ratio $\xi_{\mathrm{odlro}} / \xi_{\mathrm{phase}}$ is shown at the two levels of approximation, namely, mean field (dashed line) and $t$-matrix (full line).}
\label{Figure-7}
\end{center} 
\end{figure} 

\color{red}Figure~\ref{Figure-7} \color{black} shows the coupling dependence of the length $\xi_{\mathrm{odlro}}$ across the unitary regime at zero temperature, 
as obtained from the relation (\ref{asymptotic-regime-1}).
This length is calculated using the thermodynamic parameters $\Delta$ and $\mu$ obtained within either the mean-field or the $t$-matrix approach.
In both cases, the length $\xi_{\mathrm{odlro}}$ associated with the two-particle reduced density matrix is compared with the healing length $\xi_{\mathrm{phase}}$ associated with the static (zero-frequency) pair-pair correlation function.
It is rather remarkable how these two lengths, obtained independently in different ways, essentially coincides with each other over the whole coupling range shown in the figure.
This feature is further evidenced in the inset, where the coupling dependence of the ratio $\xi_{\mathrm{odlro}} / \xi_{\mathrm{phase}}$ is reported.
This finding confirms our expectation that a \emph{single\/} (inter-pair) coherence length can be identified along the whole BCS-BEC crossover.
Note, in particular, that, irrespective of the adopted approximation, the coupling dependence of $\xi_{\mathrm{odlro}}$ (like that of $\xi_{\mathrm{phase}}$ \cite{PS-1996}) has a characteristic 
\emph{minimum\/} close to unitarity on the BEC side, where it becomes comparable with the size $k_{F}^{-1}$ of the inter-particle distance.
Both features were regarded as distinctive properties of the coupling dependence of $\xi_{\mathrm{phase}}$ throughout the BCS-BEC crossover \cite{PS-1996}, and have recently been exploited also experimentally to highlight the proximity to this crossover \cite{Jarillo-Herrero-2021,Deutsher-2021}.

\begin{figure}[t]
\begin{center}
\includegraphics[width=6.7cm,angle=0]{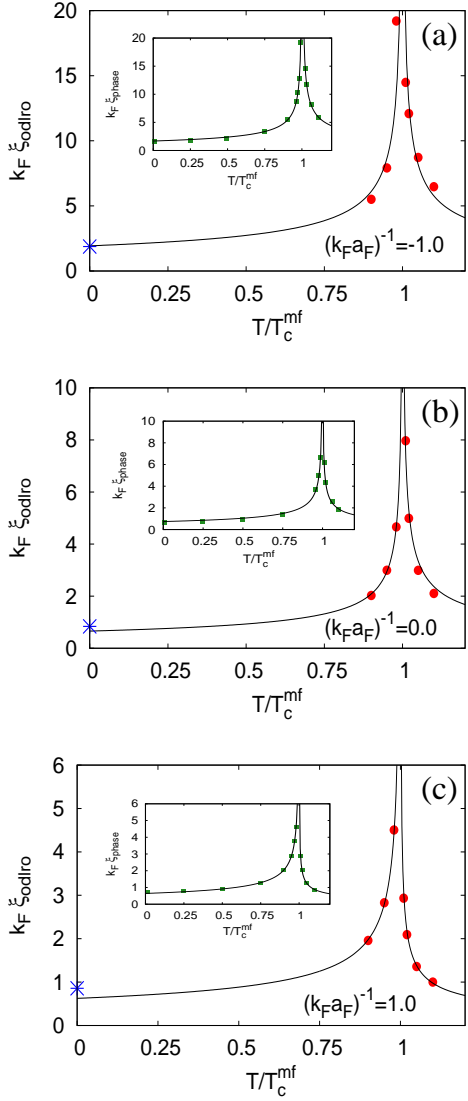}
\caption{(Color online) The temperature dependence of $\xi_{\mathrm{odlro}}$ close to $T_{c}$ (dots), obtained with mean-field thermodynamic parameters, is shown for three couplings: 
                                    $(k_{F}a_{F})^{-1}$ = -1.0 (a), 0.0 (b), and 1.0 (c). 
                                    In each case, the value of $\xi_{\mathrm{odlro}}$ at zero temperature taken from \color{red}Fig.~\ref{Figure-7} \color{black} is also shown for comparison (stars).
                                    In all panels, the solid lines represent fits of the type $k_{F} \xi_{\mathrm{fit}} / \sqrt{|1-T/T_{c}^{\mathrm{mf}}|}$ where the fitting parameter $\xi_{\mathrm{fit}}$ takes different values below and above $T_{c}^{\mathrm{mf}}$, 
                                    showing the expected temperature behavior with a mean-field critical exponent on both sides of the critical temperature $T_{c}^{\mathrm{mf}}$.
                                    The insets show the behavior of the healing length $\xi_{\mathrm{phase}}$ obtained from Ref.~\cite{PS-2014} (squares) with the corresponding fits (solid lines).}
\label{Figure-8}
\end{center} 
\end{figure} 

\color{red}Figure~\ref{Figure-8} \color{black} shows the temperature dependence of the length $\xi_{\mathrm{odlro}}$ for three distinctive couplings across the unitary regime (dots), 
in the high-temperature range $T \lesssim T_{c}$ where the expression (\ref{asymptotic-regime-3}) holds, as well as in the higher temperature range $T \gtrsim T_{c}$ where the expression 
(\ref{asymptotic-regime-4}) instead holds.
For simplicity, the thermodynamic parameters are here taken at the mean-field level.
To connect with the low-temperature results of \color{red}Fig.~\ref{Figure-7}\color{black}, fits of the type $k_{F} \xi_{\mathrm{fit}} / \sqrt{|1-T/T_{c}^{\mathrm{mf}}|}$ (solid lines) are also made through the numerical data of \color{red}Fig.~\ref{Figure-8}\color{black},
with the parameter $\xi_{\mathrm{fit}}$ taking different values below and above $T_{c}^{\mathrm{mf}}$.
In particular, below $T_{c}^{\mathrm{mf}}$ the fitting parameter $\xi_{\mathrm{fit}}$ corresponds to the value of $\xi_{\mathrm{odlro}}$ extrapolated in this way from high down to zero temperature.
For the three couplings considered in the figure, we obtain $k_{F} \xi_{\mathrm{fit}} = (1.93,0.66,0.62)$ when $(k_{F}a_{F})^{-1} = (-1.0, 0.0,1.0)$, respectively. 
It is rewarding that the extrapolated values of $k_{F} \xi_{\mathrm{odlro}}(T \rightarrow 0)$ obtained from the above fits about coincide with values of $k_{F} \xi_{\mathrm{odlro}}(T=0) = (1.88,0.84,0.86)$ at zero temperature extracted 
from \color{red}Fig.~\ref{Figure-7} \color{black} (stars).
In each panel of \color{red}Fig.~\ref{Figure-8}\color{black}, the inset shows the corresponding behavior of the healing length $\xi_{\mathrm{phase}}$ obtained from Ref.~\cite{PS-2014} (squares) with the associated temperature fit (solid line). 
From this comparison we may conclude that, not only $\xi_{\mathrm{odlro}}$ and $\xi_{\mathrm{phase}}$ (about) coincide with each other at zero temperature for all couplings as shown in \color{red}Fig.~\ref{Figure-7}\color{black},
but these two lengths also share a similar temperature dependence for given coupling (whenever it is possible to extract $\xi_{\mathrm{odlro}}$ from the relevant spatial profiles of the projected density matrix).

\begin{figure}[t]
\begin{center}
\includegraphics[width=8.5cm,angle=0]{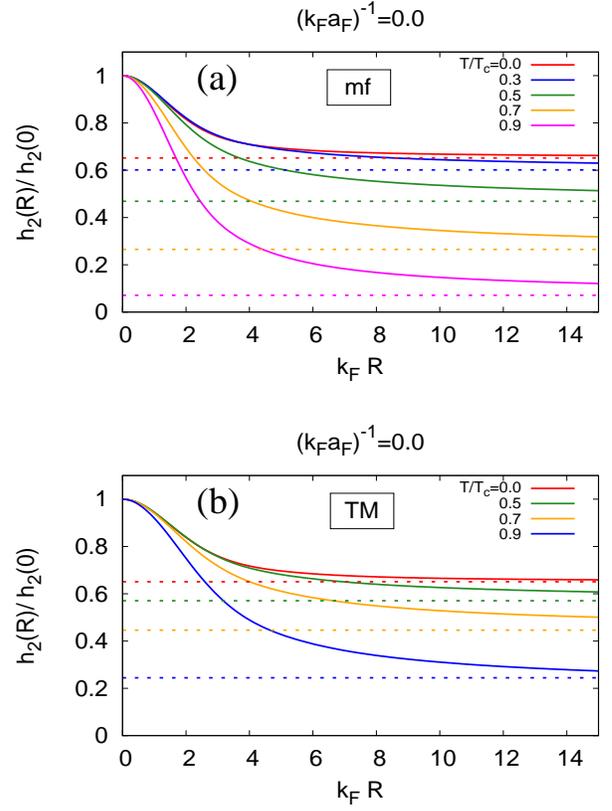}
\caption{(Color online) The radial profile of the projected density matrix $h_{2}(R) = n_{0}(T) + \delta h (R)$ (in units of its value at $R=0$) is shown at unitarity for various temperatures (full lines), 
                                     with the thermodynamic parameters calculated within mean field (mf, top panel) and the $t$-matrix approach (TM, bottom panel).
                                     In all cases, the broken lines represent the values of the condensate fraction $n_0(T)$ reached asymptotically by $h_{2}(R)$ at the given temperature.}
\label{Figure-9}
\end{center} 
\end{figure} 

Although it is not possible to determine in a meaningful way the length scale $\xi_{\mathrm{odlro}}$ from the spatial profiles of the projected density matrix $\delta h (R)$, 
in between the low- and the high-temperature regimes of the superfluid phase, even in this intermediate temperature range it is always possible to examine the \emph{overall\/} spatial dependence of the \emph{full\/} projected density matrix $h_{2}(R) = n_{0}(T) + \delta h(R)$, obtained by adding to the expression (\ref{projected-density-matrix}) for $\delta h(R)$ the value of the condensate density $n_{0}(T)$ at given temperature and coupling.
Quite generally, these profiles are obtained by setting $\boldsymbol{\rho} = \boldsymbol{\rho}'$ and integrating over $\boldsymbol{\rho}$ on the same footing, not only in the fluctuations contribution to $h_{2}$ like in Eq.~(\ref{projected-density-matrix}) for $\delta h (R)$, but also in the anomalous contribution to $h_{2}$ given by Eq.~(\ref{first-term-h_2-general}).
The results obtained in this way at unitarity are shown in \color{red}Fig.~\ref{Figure-9} \color{black} for various temperatures in the superfluid phase (full lines), where the thermodynamic parameters $\Delta$ and $\mu$ obtained either within the mean-field (top panel) or the $t$-matrix approach (bottom panel).
These plots help one to visualize how the convergence of $h_{2}(R)$ to its asymptotic value $n_{0}(T)$ (broken lines) occurs in practice, and, in particular, how it progressively crosses over from a power-law behavior at low temperature to an exponential decay at high temperature.

\begin{figure}[t]
\begin{center}
\includegraphics[width=8.5cm,angle=0]{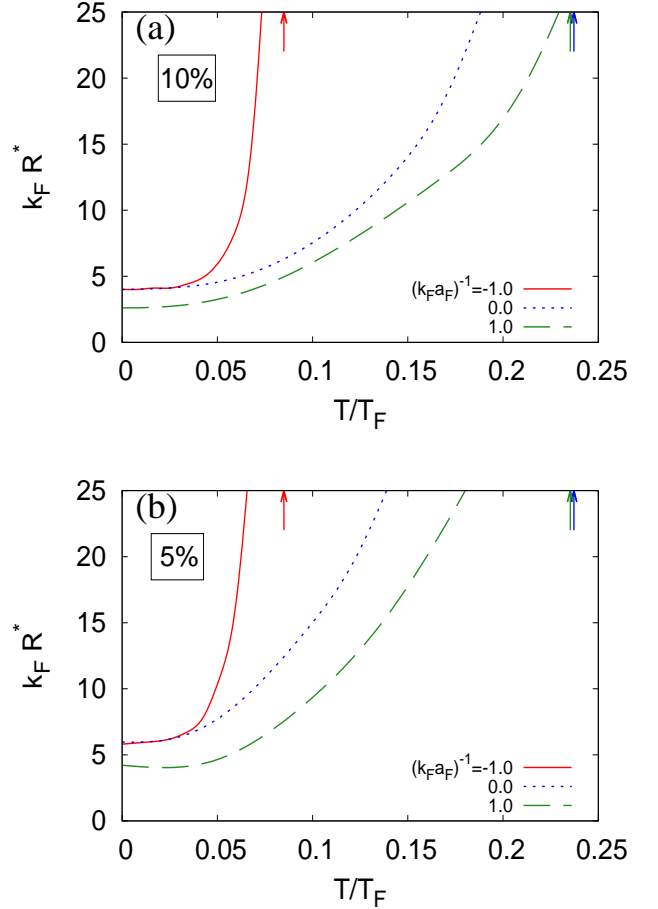}
\caption{(Color online) The temperature dependence of the distance $R^{*}$ (in units of the inverse of the Fermi wave vector $k_{F}$), at which $h_{2}(R)$ has reached its asymptotic value 
                                     $n_{0}(T)$ for given coupling with a relative error of 
                                     (a) $10 \%$ and (b) $5 \%$, is shown for three representative couplings.
                                     The calculations are here done within the $t$-matrix approach.
                                     In both panels, the arrows indicate the corresponding value of the critical temperature for given coupling.}
\label{Figure-10}
\end{center} 
\end{figure} 

In this context, it would also be of practical importance (especially when considering finite-size systems, like ultra-cold Fermi gases or nuclei \cite{Physics-Reports-2018}) to identify the distance $R^{*}$, at which the function $h_{2}(R)$ reaches its asymptotic value $n_{0}(T)$ \emph{within a given relative error\/}, for given coupling and temperature. 
The values of $R^{*}$ identified in this way are shown in \color{red}Fig.~\ref{Figure-10}\color{black}, where two relative errors of $10 \%$ and $5 \%$ are specifically considered. 
In this case the calculations were done only with the $t$-matrix approach, since it is expected to be more reliable than the mean-field approach.
For instance, for the $5 \%$ relative error considered in \color{red}Fig.~\ref{Figure-10}(b)\color{black}, when $T \simeq 0.1 T_{F}$ the value of $R^{*}$ at unitarity is about $15 \, k_{F}^{-1}$.

It is further interesting to compare the values of $R^{*}$ obtained in this way with the size of the cloud of a Fermi gas embedded in a harmonic trapping potential at zero temperature, for which the Thomas-Fermi radius $R_{\mathrm{TF}}$ of the non-interacting case represents an upper bound.
This radius is given by $k_{F}(0) R_{\mathrm{TF}} = (48 N)^{1/3}$, where $k_{F}(0) = (6 \pi^{2} n(0))^{1/3}$ is the Fermi wave-vector corresponding to the density $n(0)$ at the center of the cloud and $N$ is the total number of fermions.
For typical experimental values $N \approx 10^{4} -10^{5}$ one obtains correspondingly $R_{\mathrm{TF}} \approx (100 - 200) k_{F}(0)^{-1}$, which is one order of magnitude larger than the representative value  
$R^{*} \simeq 15 \, k_{F}^{-1}$ indicated above.

This result may also serve to provide a pragmatic answer to the question asked for trapped gases in Ref.~\cite{Leggett-2006}, where a concern was raised about utilizing the definition of the order parameter 
based on ODLRO, to the extent that, strictly speaking, for finite-size systems the limit $R \rightarrow \infty$ cannot be taken in a sensible way.
This is because, whenever the distance $R^{*}$ can be considered to be much smaller than the typical size of the cloud, in practice the occurrence of asymptotic correlations in the two-particle reduced density matrix should be taken for granted, at least within a given default uncertainty.

\begin{figure}[t]
\begin{center}
\includegraphics[width=8.5cm,angle=0]{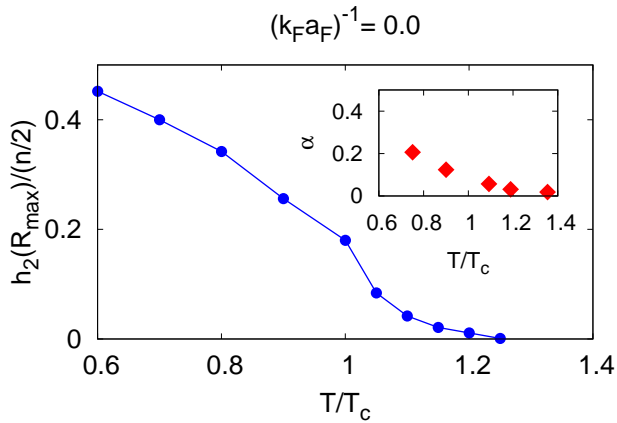}
\caption{(Color online) The projected density matrix $h_{2}(R=R_{\mathrm{max}})$ for the finite value $R_{\mathrm{max}} = 10 k_{F}^{-1}$, calculated at unitarity within the $t$-matrix approach like in \color{red}Fig.~\ref{Figure-9}\color{black}, 
                                    is shown as a function of temperature across the critical temperature $T_{c}$.
                                    The inset reports the values of the condensate fraction $\alpha$ taken from Fig.~1 of Ref.~\cite{Bulgac-2020}, as obtained by QMC calculations on systems of finite size.}
\label{Figure-11}
\end{center} 
\end{figure} 

Finally, the results for $h_{2}(R) = n_{0}(T) + \delta h (R)$, which were reported in \color{red}Fig.~\ref{Figure-9} \color{black} below $T_{c}$, can also be extended above $T_{c}$ because the term containing $\Gamma_{11}$ in Eq.~(\ref{explicit-form-delta-h_2})
survives in the normal phase even though $n_{0}(T \ge T_{c})$ vanishes therein.
This property enables us to examine the behavior of $h_{2}(R)$ across the critical temperature when it is calculated for a \emph{finite\/} value $R_{\mathrm{max}}$ of the radial variable $R$. 
This is shown in \color{red}Fig.~\ref{Figure-11} \color{black} at unitarity for the value $R_{\mathrm{max}} = 10 k_{F}^{-1}$, where $h_{2}(R=R_{\mathrm{max}})$ not only remains finite at $T_{c}$ but also shows a ``convex'' behavior for $T \gtrsim T_{c}$.
In this context, it is interesting to mention that a similar convex behavior across $T_{c}$ was reported in Ref.~\cite{Bulgac-2020} for the condensate density obtained by QMC calculations performed on systems of finite size, as shown in the inset
of \color{red}Fig.~\ref{Figure-11}\color{black}.


\vspace{-0.4cm}
\section{Concluding remarks}
\label{sec:conclusions}
\vspace{-0.2cm}

In this article, we have considered the two-particle reduced density matrix, which has long been identified as the central quantity for establishing the superfluid properties of a fermionic system \cite{Yang-1962}, and calculated it numerically in \emph{all\/} of its aspects
throughout the BCS-BEC crossover.
To this end, we have relied on a diagrammatic approach that includes beyond-mean-field pairing fluctuations, which has proved sufficient to describe the relevant features of the BCS-BEC crossover \cite{Physics-Reports-2018}.
The novelty here is that we have not only been concerned with the condensate density, which is extracted from the two-particle reduced density matrix in the limit when either Cooper pairs (in the BCS limit) or bosonic dimers (in the BEC limit) recede far apart from each other, 
but we have also determined the spatial range past which the asymptotic correlations associated with ODLRO are effectively established,
an information which can be of practical importance especially for finite-size systems \cite{Leggett-2006}.

Recall in this context that, for a Bose-Einstein condensate, spatial coherence is measured by means of the fringe visibility in interferometric experiments, which shows an algebraic decay for large separations as expected according to the 
off-diagonal elements of the one-particle reduced density matrix \cite{Bloch-2000}.
For a Fermi gas, on the other hand, spatial coherence emerges only via the two-particle reduced density matrix, which requires more sophisticated interferometric techniques to be detected \cite{Carusotto-2005}. 

Whenever possible on physical grounds \cite{BKV-2005}, from our numerical calculations we have also extracted the
coupling- and temperature-dependent length $\xi_{\mathrm{odlro}}$ specifically associated with ODLRO, and found that it (about) coincides with the
length $\xi_{\mathrm{phase}}$ associated instead with the static limit of the pair-pair correlation function considered some time ago in the context of the BCS-BEC crossover \cite{PS-1996,PS-2014}.
In particular, at low temperature both lengths as function of coupling show a characteristic minimum where they reach the value of the inter-particle distance.
This feature has been recently utilized in experiments on condensed-matter samples \cite{Jarillo-Herrero-2021,Deutsher-2021}, to identify what would correspond to the unitary regime of the BCS-BEC crossover
with ultra-cold Fermi gases.

In the present article, the BCS-BEC crossover has been exploited as a theoretical tool to make the entities, which recede far apart from each other in the ODLRO protocol, to evolve in a continuous fashion from Cooper pairs to bosonic dimers.
In the process, we have verified that the \emph{two-particle\/} reduced density matrix for the constituent fermions effectively evolves, when passing from the BCS to the BEC limits, into the \emph{one-particle\/} reduced density matrix for (composite) bosons made up of tightly-bound fermion pairs.
With the experimental methodologies nowadays available for ultra-cold Fermi gases \cite{Ketterle-Zwierlein-2007}, it should then be possible to turn this ``gedankenexperiment'' into a real experiment by measuring the frequency and wave-vector dependence of the pair-pair correlation function, from which the length $\xi_{\mathrm{phase}}$ could be obtained in the limit of zero-frequency and the length $\xi_{\mathrm{odlro}}$
by an averaging over frequencies, once the wave-vector dependence has been turned into spatial profiles.


\begin{center}
\begin{small}
{\bf ACKNOWLEDGMENTS}
\end{small}
\end{center}
\vspace{-0.2cm}

Partial financial support from the Italian MIUR under Project PRIN2017 (20172H2SC4) is acknowledged.
This article is dedicated to Prof.~Chen-Ning Yang who first proposed the concept of ODLRO.

\appendix   

 
\vspace{-0.4cm}
\section{Analytic results for the fermionic \\ one-particle reduced density matrix \\ at $T=0$ within the mean-field approximation} 
\label{sec:appendix-A}
\vspace{-0.2cm}

In this Appendix, we consider the expression (\ref{G_11-mean-field}) of the fermionic one-particle reduced density matrix $\mathcal{G}_{11}^{\mathrm{mf}}(\mathbf{R},0^{-})$ at the mean-field level, and obtain analytically its asymptotic spatial width $\xi_{1}$ at any temperature throughout the BCS-BEC crossover.
It will turn out that $\xi_{1}$ (about) coincides with the Cooper pair size $\xi_{\mathrm{pair}}$ \cite{PS-1994,PS-2014}, as determined from the anomalous term $\mathcal{G}_{12}^{\mathrm{mf}}(\boldsymbol{\rho},0^{-})$ given by Eq.~(\ref{G_12-mean-field}).
The results here discussed complement the numerical analysis of Sec.~\ref{subsec:results_normal} and extend to any temperature throughout the BCS-BEC crossover the results obtained
in Ref.~\cite{Fetter-1965} at any temperature but in the weak-coupling (BCS) limit only, on the one hand, and in Ref.~\cite{Romero-2020} throughout the BCS-BEC crossover but at zero temperature only, on the other hand.

\begin{figure*}[t]
\begin{center}
\includegraphics[width=16.8cm,angle=0]{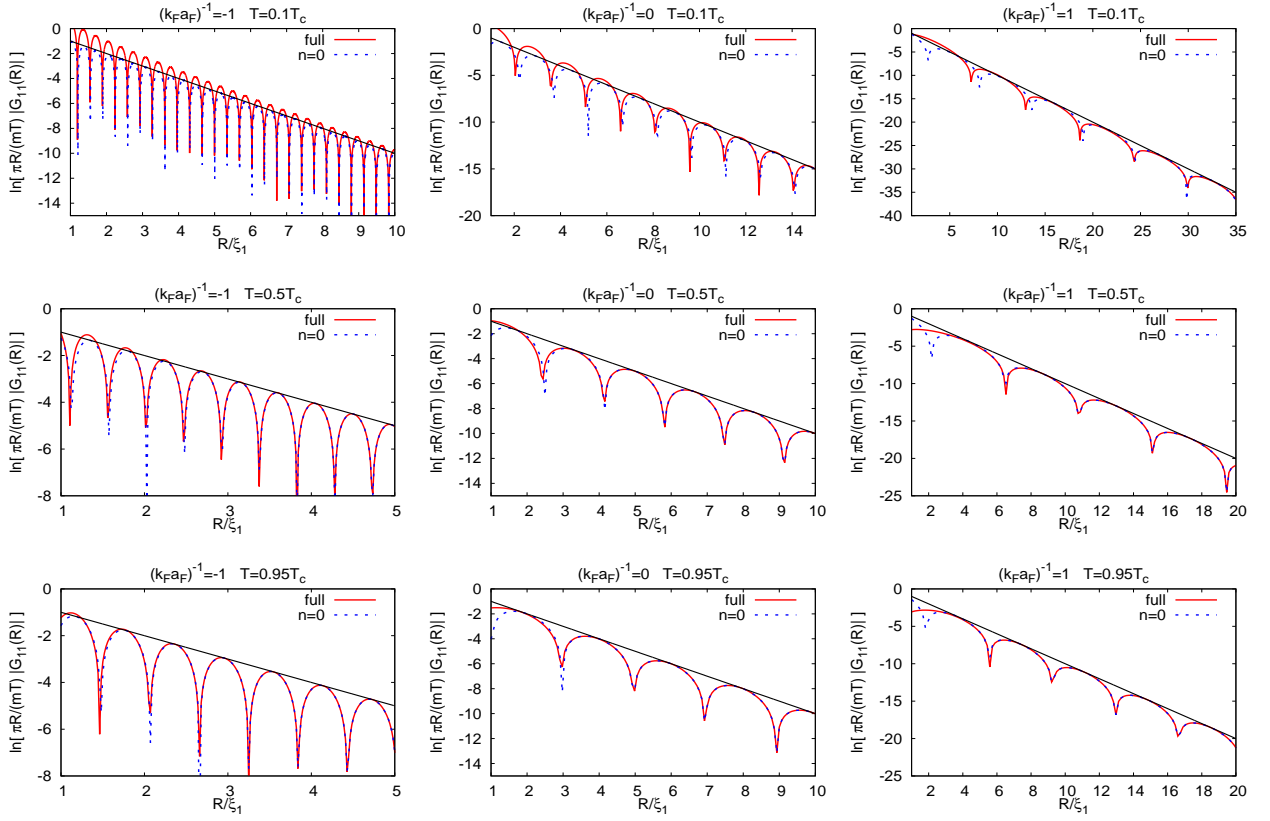}
\caption{(Color online) The function $\ln \! \left[ \frac{\pi R}{m T} |\mathcal{G}_{11}^{\mathrm{mf}}(R,0^{-})| \right]$, obtained from the expression (\ref{resulting-expression-final}) with $n_{\mathrm{max}} = 100$ 
                                    (full lines) and $n=0$ only (dashed lines), is shown versus $R/\xi_{1}$ for three characteristic couplings across the BCS-BEC crossover and three temperatures, from $T \simeq 0$ to 
                                    $T \simeq T_{c}$. 
                                    In each panel, the straight lines $-R/\xi_{1}$ represent the asymptotic behavior of this function (apart from a constant term), where the length $\xi_{1}$ depends on coupling and 
                                    temperature according to the expression (\ref{xi_1}).}                                   
\label{Figure-12}
\end{center} 
\end{figure*} 

We begin by rewriting the expression within braces in the second line of Eq.~(\ref{G_11-mean-field}) in the form
\begin{eqnarray}
& & u(\mathbf{k})^{2} f(E(\mathbf{k})) + v(\mathbf{k})^{2} \left[ 1 - f(E(\mathbf{k})) \right]  
\nonumber \\
& = & \frac{1}{2} - \xi(\mathbf{k}) \, \frac{2}{\beta} \, \sum_{n = 0}^{+\infty} \frac{1}{\omega_{n}^{2} + E(\mathbf{k})^{2}} \, ,
\label{rewritten-expression}
\end{eqnarray}
obtained by using the spectral representation of the Fermi function 
\begin{equation}
f(\epsilon) = \frac{1}{\beta} \, \sum_{n = -\infty}^{+\infty} \frac{e^{i \omega_{n} \eta}}{i \omega_{n} - \epsilon} 
\label{Fermi-function}
\end{equation}
in terms of fermionic Matsubara frequencies $\omega_{n}$ \cite{FW-1971}.
Once the expression (\ref{rewritten-expression}) is inserted into Eq.~(\ref{G_11-mean-field}), the constant ($\frac{1}{2}$) term therein gives rise to a Dirac delta function in the spatial variable $\mathbf{R}$, 
which we will consistently omit in the following.
Upon performing the integration over the angle between $\mathbf{k}$ and $\mathbf{R}$ in Eq.~(\ref{G_11-mean-field}) and integrating by parts the remaining radial integral in $k = |\mathbf{k}|$, 
we are left with the expression
\begin{equation}
\mathcal{G}_{11}^{\mathrm{mf}}(R,0^{-}) = I(R) + I(R)^{*}
\label{resulting-expression}
\end{equation}
for $R = |\mathbf{R}| > 0$, where
\begin{eqnarray}
I(R) & = & \frac{i}{2 \pi^{2} m R} \frac{d}{d R} \frac{1}{R \beta} \sum_{n=0}^{+\infty} \, \int_{0}^{\infty} \!\! dk \, k \, e^{ikR} 
\nonumber \\
& \times & \left[ \frac{1}{\omega_{n}^{2} + E(\mathbf{k})^{2}} - \frac{2 \, \xi(\mathbf{k})^{2}}{(\omega_{n}^{2} + E(\mathbf{k})^{2})^{2}} \right] \, .
\label{I-R}
\end{eqnarray}
We can follow at this point Ref.~\cite{Fetter-1965} and introduce the auxiliary function
\begin{equation}
K_{n}(R;\lambda) = - i \! \int_{0}^{\infty} \!\! dk \, k \, e^{ikR} \, \frac{1}{\omega_{n}^{2} + \lambda^{2} \xi(\mathbf{k})^{2} + \Delta^{2}} \, ,
\label{auxiliary-function}
\end{equation}
in such a way that the expression (\ref{I-R}) is rewritten in the compact form:
\begin{eqnarray}
I(R) & = & - \frac{1}{2 \pi^{2} m R} \frac{d}{d R} \frac{1}{R \beta} \sum_{n=0}^{+\infty} \, \int_{0}^{\infty} \!\! dk \, k \, e^{ikR} 
\nonumber \\
& \times & \sum_{n=0}^{+\infty} \, \frac{\partial}{\partial \lambda} \left. \left( \lambda K_{n}(R;\lambda) \right) \right|_{\lambda = 1} \, .
\label{I-R-compact}
\end{eqnarray}
To proceed further, we look for the zeros of the denominator in Eq.~(\ref{auxiliary-function}) in the complex $k$-plane and write
\begin{eqnarray}
&& \left( \frac{2 \, m}{\lambda} \right)^{2} \! \left( \omega_{n}^{2} + \lambda^{2} \xi(\mathbf{k})^{2} + \Delta^{2} \right) = [k - (q_{n} + i p_{n})] 
\nonumber \\
& \times & [k - (q_{n} - i p_{n})] [k + (q_{n} + i p_{n})] [k + (q_{n} - i p_{n})]
\nonumber 
\end{eqnarray}
where
\begin{equation}
\left\{ 
\begin{array}{c}
q_{n} =  \sqrt{ m \sqrt{ \mu^{2} + \frac{\omega_{n}^{2} + \Delta^{2}}{\lambda^{2}}} + m \mu } \\
p_{n} =  \sqrt{ m \sqrt{ \mu^{2} + \frac{\omega_{n}^{2} + \Delta^{2}}{\lambda^{2}}} - m \mu } \, .
\end{array}
\right.
\label{qn-pn}
\end{equation}
To the $k$-integral from $k=0$ up to $k=+\infty$ along the real axis in Eq.~(\ref{auxiliary-function}) we can now add a vanishing contribution coming from a large semicircle in the first quadrant of the complex $k$-plane, plus the integral along the imaginary axis from $k=+i \infty$ down to $k=0$ which is not going, however, to affect the expression (\ref{resulting-expression}) we are after.
In this way, the $k$-integral in Eq.~(\ref{auxiliary-function}) can be transformed into an integral over a closed curve that encircles the pole at $(q_{n} + i p_{n})$.
Applying Cauchy's integral formula over this closed curve then gives 
\begin{equation}
K_{n}(R;\lambda) = - \frac{i \pi m}{\lambda} \, \frac{ e^{i(q_{n}+ip_{n}) R}} {\sqrt{\omega_{n}^{2} + \Delta^{2}}} \, ,
\label{Cauchy-integral-formula}
\end{equation}
which can be utilized in Eq.~(\ref{I-R-compact}) to obtain
\begin{equation}
I(R) = - \frac{m}{2 \pi R} \frac{1}{\beta} \sum_{n=0}^{+\infty} e^{i(q_{n}+ip_{n}) R} \, .
\label{final-expression-I-R} 
\end{equation}
This yields eventually the desired result [cf. Eq.~(\ref{resulting-expression})]
\begin{equation}
\mathcal{G}_{11}^{\mathrm{mf}}(R,0^{-}) \, = \, - \frac{m}{\pi R} \frac{1}{\beta} \sum_{n=0}^{+\infty} e^{-p_{n} R} \cos(q_{n} R) \, .
\label{resulting-expression-final}
\end{equation}
From this expression one gets that the leading contribution to $\mathcal{G}_{11}^{\mathrm{mf}}(R,0^{-})$ for large $R$ stems from the smallest value of $p_{n}$, 
which corresponds to the term with $n=0$ according to Eq.~(\ref{qn-pn}).
This identifies the \emph{characteristic length\/} $\xi_{1}$ of the fermionic one-particle reduced density matrix as follows
\begin{equation}
\xi_{1} = \frac{1}{p_{n=0}} = \left[ m \left( \sqrt{ \mu^{2} + \Delta^{2} + \frac{\pi^{2}}{\beta^{2}} } - \mu \right) \right]^{-1/2} \, ,
\label{xi_1}
\end{equation}
which holds at the mean-field level for any coupling and temperature.

The coupling and temperature dependence of $\xi_{1}$ was reported in Sec.~\ref{sec:numerical_results}-B, where it was also compared with the corresponding behavior of the Cooper pair size.
Here, we instead show the full spatial profile of the expression (\ref{resulting-expression-final}) (for which we have found it sufficient to extend the sum over $n$ up to $n_{\mathrm{max}} = 100$,
in order to obtain good convergence over the whole considered spatial range) and compare it with the spatial profile obtained by the $n=0$ term only.
This is done in \color{red}Fig.~\ref{Figure-12}\color{black}, where the function $\ln \! \left[ \frac{\pi R}{m T} |\mathcal{G}_{11}^{\mathrm{mf}}(R,0^{-})| \right]$ is shown versus $R/\xi_{1}$ for a choice of temperatures and couplings in both cases, namely, with $n_{\mathrm{max}} = 100$ (full lines) and $n=0$ only (dashed lines).
In each panel, this comparison evidences that retaining the $n=0$ term only represents a good approximation to the full function on the BCS side of the crossover, while the comparison becomes slightly worse on the BEC side.
In addition, in each panel the straight line $-R/\xi_{1}$ corresponds to the asymptotic damping envelope of the expression (\ref{resulting-expression-final}) and evidences the role played by the length scale (\ref{xi_1}) in that expression.
For the lowest temperature considered in \color{red}Fig.~\ref{Figure-12}\color{black}, these plots reflect the same behavior reported in Ref.~\cite{Romero-2020} across the BCS-BEC crossover albeit at zero temperature only.

Finally, the expression (\ref{G_11-mean-field}) for $\mathcal{G}_{11}^{\mathrm{mf}}(\mathbf{R},0^{-})$ can be readily calculated in the normal phase for temperatures much larger than $T_{c}$ (such that $\mu / k_{B}T \rightarrow - \infty$), where it recovers the classical (Boltzmann) result for non-interacting particles
\begin{equation}
\mathcal{G}_{0}(R|T \gg T_{F}) \, \simeq \, e^{\beta \mu} \, \frac{ e^{ -\frac{\pi R^{2}}{\lambda_{T}^{2}} } }{ \lambda_{T}^{3} }
\label{Boltzmann-limit}
\end{equation}
with $\lambda_{T} = \sqrt{\frac{2 \pi}{m k_{B} T}}$ the thermal wavelength.
In this limit, we can identify $\xi_{1} = \lambda_{T}/\sqrt{\pi}$, to be compared with the corresponding result  $\xi_{\mathrm{pair}} = \lambda_{T}/(2 \sqrt{\pi})$ for the Cooper pair size \cite{PS-2014}. 
This yields $\xi_{1} / \xi_{\mathrm{pair}} = 2$, in line with the results shown in the insets of \color{red}Figs.~\ref{Figure-6}(a) \color{black} and \color{red}\ref{Figure-6}(b)\color{black}.


\vspace{-0.4cm}
\section{Analytic results for the bosonic \\ one-particle reduced density matrix \\ at $T=0$ within the Bogoliubov approximation} 
\label{sec:appendix-B}
\vspace{-0.2cm}

In this Appendix, we consider the bosonic one-particle reduced density matrix obtained from the Bogoliubov approximation (\ref{bosonic-normal-propagator}) for the normal single-particle Green's function $\mathcal{G}^{'}_{B}$, which is associated with the particles out of the condensate.
In particular, at zero temperature the term containing the Bose function (cf. Eq.~(\ref{dominant-term-BEC-limit-final})) drops out and one obtains: 
\begin{eqnarray}
& - & \frac{\mathcal{G}^{'}_{B}(R)}{n'_{B}} = \frac{1}{4 \pi^{2} n'_{B} R} \!\! \int_{0}^{+\infty} \!\!\!\!\!\!\!\! dQ \, \sin(Q R) \, Q \! \left[ \frac{ \frac{Q^{2}}{2 \, m_{B}} + \mu_{B} }{ E_{B}(Q) } -  1 \right]
\nonumber \\
& = & \frac{1}{4 \pi^{2}  n'_{B} \, \tilde{\xi}_{B}^{3}} \, \frac{\tilde{\xi}_{B}}{R} \!
\int_{0}^{+\infty} \!\!\!\!\!\!\!\! dx \, \sin \left( \! \frac{R}{\tilde{\xi}_{B}} x \! \right) x \! \left[ \frac{1 + x^{2} }{ \sqrt{x^{4}+2x^{2}} } -  1 \right] 
\label{bosonic-OPRDM-zero-T}
\end{eqnarray}

\begin{figure}[t]
\begin{center}
\includegraphics[width=7.5cm,angle=0]{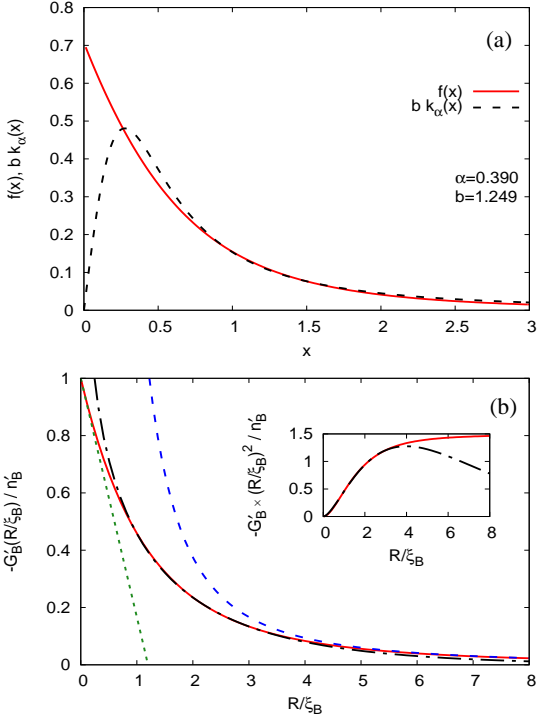}
\caption{(Color online) (a) Comparison between the functions $f(x)$ of Eq.~(\ref{function-f}) (full line) and $b \,k_{\alpha}(x)$ of Eq.~(\ref{function-k}) (dashed line), with the reported values of $b$ and $\alpha$. 
                                    (b) Comparison between the spatial profiles of $-\mathcal{G}^{'}_{B}(R)/n'_{B}$ of Eq.~(\ref{bosonic-OPRDM-zero-T}) (full line) and $b K_{0}(\alpha R/\tilde{\xi}_{B})/(4 \pi^{2} n'_{B} \, \tilde{\xi}_{B}^{3})$ of 
                                          Eqs.~(\ref{modified-K_0})-(\ref{function-k}) (dashed-dotted line), with the same values of $b$ and $\alpha$ reported in panel (a).
                                          Also shown are the linear approximation to $-\mathcal{G}^{'}_{B}(R)/n'_{B}$ for $R \lesssim \tilde{\xi}_{B}$ (dotted line) and the asymptotic $(R/\tilde{\xi}_{B})^{-2}$ behavior of $-\mathcal{G}^{'}_{B}(R)/n'_{B}$ for $R \gg \tilde{\xi}_{B}$
                                          (dashed line).
                                    The inset shows $-\mathcal{G}^{'}_{B}(R)/n'_{B}$ (full line) and $b K_{0}(\alpha R/\tilde{\xi}_{B})/(4 \pi^{2} n'_{B} \, \tilde{\xi}_{B}^{3})$ (dashed-dotted line), both multiplied by $(R/\tilde{\xi}_{B})^{2}$ to evidence their different asymptotic 
                                    behavior for $R \gg \tilde{\xi}_{B}$.}
\label{Figure-13}
\end{center} 
\end{figure} 

\noindent
with $Q=|\mathbf{Q}|$.
Here, $n'_{B}$ is the non-condensate density, $m_{B}$ the particle mass, $\mu_{B} = \frac{4 \pi a_{B}}{m_{B}} n_{B}$ the chemical potential with scattering length $a_{B}$ and particle density $n_{B}$, 
and $\tilde{\xi}_{B}= (2 m_{B} \mu_{B})^{-1/2}$ the healing length in the present context (with the suffix $B$ referring to bosonic quantities) \cite{footnote-3}.    
The expression within brackets in the second line of Eq.~(\ref{bosonic-OPRDM-zero-T}), with the rescaled integration variable $x = Q \, \tilde{\xi}_{B}$ in the place of the original variable $Q$, reduces to  $(\sqrt{2} x)^{-1}$ for $x \ll 1$ and to $(2 x^{4})^{-1}$ for $x \gg 1$.
The singular small-$x$ behavior is responsible for the asymptotic $R^{-2}$ tail of the Fourier transform $\mathcal{G}^{'}_{B}(R)$ for $R \gg \tilde{\xi}_{B}$, while the large-$x$ behavior is related to the Tan contact $C$ \cite{Tan-2008-I}-\cite{Tan-2008-III} 
and gives rise to a cusp in $\mathcal{G}^{'}_{B}(R)$ at $R=0$ \cite{footnote-4}.

We can also determine the behavior of $\mathcal{G}^{'}_{B}(R)$ for \emph{intermediate} values of $R$ lying between $\tilde{\xi}_{B} \lesssim R$ and $R \gg \tilde{\xi}_{B}$, which should account in practice for most part of the spatial profile of $\mathcal{G}^{'}_{B}(R)$.
To this end, we exploit the formal similarity between the function occurring in the second line of Eq.~(\ref{bosonic-OPRDM-zero-T}), which we rewrite in the form  
\begin{eqnarray}
\mathcal{F} \! \left( \! \frac{R}{\tilde{\xi}_{B}} \! \right) & = & \frac{\tilde{\xi}_{B}}{R} \! \int_{0}^{+\infty} \!\!\!\!\!\!\! dx \, \sin \left( \! \frac{R}{\tilde{\xi}_{B}} \, x \! \right) f(x) 
\label{function-F} \\
f(x) & = & x \! \left[ \frac{1 + x^{2} }{ \sqrt{x^{4}+2x^{2}} } -  1 \right] \, ,
\label{function-f}
\end{eqnarray}
and the integral representation of the modified Bessel function of zero order \cite{AS-NBS-1972}, which can be cast in the form
\begin{eqnarray}
K_{0} \! \left( \! \alpha \frac{R}{\tilde{\xi}_{B}} \! \right) & = & \frac{\tilde{\xi}_{B}}{R} \int_{0}^{+\infty} \!\!\!\!\!\!\! dx \, \sin \left( \! \frac{R}{\tilde{\xi}_{B}} \, x \! \right) \, k_{\alpha}(x) 
\label{modified-K_0} \\
k_{\alpha}(x) & = & \frac{1}{\alpha^{2}} \, \frac{ \frac{x}{\alpha} }{ \left( 1 +  \frac{x^{2}}{\alpha^{2}} \right)^{3/2} } \, .
\label{function-k}
\end{eqnarray}
We thus multiply the function $k_{\alpha}(x)$ by an overall factor $b$, and vary $b$ and $\alpha$ so as to optimize the comparison between $k_{\alpha}(x)$ and $f(x)$ of Eq.~(\ref{function-f}), 
with emphasis on an extended interval of intermediate values of $x$ centered about $x=1$. 
This is because we expect this interval to be relevant for the corresponding interval of intermediate values of $R/\tilde{\xi}_{B}$, where we would like to optimize the comparison between the corresponding (sine) Fourier transforms $b K_{0}(\alpha R/\tilde{\xi}_{B})$ 
and $\mathcal{F}(R/\tilde{\xi}_{B})$.

\color{red}Figure~\ref{Figure-13}(a) \color{black} compares the functions $f(x)$ of Eq.~(\ref{function-f}) (full line) and $b \,k_{\alpha}(x)$ of Eq.~(\ref{function-k}) (dashed line), where the fitting parameters $b$ and $\alpha$ are suitably chosen to optimize the comparison.
With the values of $b$ and $\alpha$ determined in this way, \color{red}Fig.~\ref{Figure-13}(b) \color{black} then compares the spatial profiles of the functions $- \mathcal{G}^{'}_{B}(R)/n'_{B}$ and $b K_{0}(\alpha R/\tilde{\xi}_{B})/(4 \pi^{2}  n'_{B} \, \tilde{\xi}_{B}^{3})$ over an extended interval of $R$ spanning several times $\tilde{\xi}_{B}$.
Note how $b K_{0}(\alpha R/\tilde{\xi}_{B})/(4 \pi^{2}  n'_{B} \, \tilde{\xi}_{B}^{3})$ well approximates $- \mathcal{G}^{'}_{B}(R)/n'_{B}$ over the extended interval $\tilde{\xi}_{B} \lesssim R \lesssim 5 \, \tilde{\xi}_{B}$ of intermediate values of $R$, 
with $\mathcal{G}^{'}_{B}(R \approx 5 \tilde{\xi}_{B}) \approx \mathcal{G}^{'}_{B}(R =0)/20$.
\color{red}Figure~\ref{Figure-13}(b) \color{black} also reports for comparison the linear approximation to $-\mathcal{G}^{'}_{B}(R)/n'_{B}$ for $R \lesssim \tilde{\xi}_{B}$ and the asymptotic $(R/\tilde{\xi}_{B})^{-2}$ tail for $R \gg \tilde{\xi}_{B}$.
Although in these outer ranges of $R$ the function $b K_{0}(\alpha R/\tilde{\xi}_{B})/(4 \pi^{2}  n'_{B} \, \tilde{\xi}_{B}^{3})$ fails to well approximate $-\mathcal{G}^{'}_{B}(R)/n'_{B}$, the resulting discrepancies appear irrelevant to our main conclusion
that $(-4 \pi^{2} \, \tilde{\xi}_{B}^{3}) \mathcal{G}^{'}_{B}(R)$ converges to zero essentially like $b \, \sqrt{\frac{\pi \tilde{\xi}_{B}}{2 \alpha R}}e^{- \alpha R/\tilde{\xi}_{B}}$ \cite{AS-NBS-1972} and that this convergence is exhausted when $R$ has reached 
about a few times $\tilde{\xi}_{B}$.

This piece of physical information is relevant to the numerical analysis made in Sec.~\ref{subsec:results_fluctuations} when dealing with a gas of superfluid fermions undergoing the BCS-BEC crossover. 
However, a different (and, in practice, more efficient) systematic procedure was there utilized to identify (whenever possible) the length scale $\xi_{\mathrm{odlro}}$ over which the projected density matrix $\delta h(R)$ of Eq.~(\ref{projected-density-matrix}), obtained from the expression (\ref{explicit-form-delta-h_2}) of the contribution by pairing fluctuations to the two-particle reduced density matrix, converges to zero.


\vspace{-0.4cm}
\section{Asymptotic behavior of \\ the two-particle reduced density matrix \\ in the BCS limit at zero temperature} 
\label{sec:appendix-C}
\vspace{-0.2cm}

In this Appendix, we extend to the weak-coupling (BCS) limit of the BCS-BEC crossover the estimate for the leading asymptotic spatial behavior of the projected density matrix $\delta h(R)$ given by Eqs.~(\ref{projected-density-matrix}) and (\ref{explicit-form-delta-h_2}), that was considered in Sec.~\ref{subsec:fluctuations-contribution-BEC} in the opposite strong-coupling (BEC) limit.
Here, our analysis will be limited to zero temperature, for which the analytic results of Ref.~\cite{MPS-1998} in terms of elliptic integrals will be exploited.

In the BCS limit, whereby $\Delta/\mu << 1$ and $\mu \simeq E_{F}$ \cite{Physics-Reports-2018}, the following leading contribution to $\delta h(R)$ is obtained from Eq.~(\ref{explicit-form-delta-h_2}):
\begin{eqnarray}
\delta h(R) & \simeq & \int \!\!\! \frac{d \mathbf{Q}}{(2 \pi)^{3}} \, e^{i \mathbf{Q} \cdot \mathbf{R}} \, \frac{1}{\beta} \sum_{\nu} e^{i \Omega_{\nu} \eta} \, \Gamma_{11}(\mathbf{Q},\Omega_{\nu}) 
\nonumber \\
& \times &  \int \!\! d\boldsymbol{\rho} \,\, \tilde{\Pi}_{11}(\boldsymbol{\rho};\mathbf{Q},\Omega_{\nu}) \, \tilde{\Pi}_{11}(-\boldsymbol{\rho};\mathbf{Q},\Omega_{\nu}) \, .
\label{dominant-term-BCS-limit}
\end{eqnarray}
Although this result is  formally similar to Eq.~(\ref{dominant-term-BEC-limit}) obtained in the BEC limit, the two factors of the $\mathbf{Q}$-integrand in Eq.~(\ref{dominant-term-BCS-limit}) now acquire values specific to the BCS limit.
In particular, we are interested in their leading small-$Q$ behavior (where $Q = |\mathbf{Q}|$), since this accounts for the leading large-$R$ behavior of the Fourier transform in Eq.~(\ref{dominant-term-BCS-limit}) \cite{Lighthill-1959}.
With the help of the analytic results of Ref.~\cite{MPS-1998}, we obtain accordingly:
\begin{eqnarray}
&& \int \!\! d\boldsymbol{\rho} \,\, \tilde{\Pi}_{11}(\boldsymbol{\rho};\mathbf{Q},\Omega_{\nu}) \, \tilde{\Pi}_{11}(-\boldsymbol{\rho};\mathbf{Q},\Omega_{\nu}) 
\nonumber \\
& = & \int \!\!\! \frac{d \mathbf{k}}{(2 \pi)^{3}} \left[ \frac{1}{\beta} \, \sum_{n} \mathcal{G}_{11}^{\mathrm{mf}}(\mathbf{k} + \mathbf{Q},\omega_{n}+\Omega_{\nu}) \, \mathcal{G}_{11}^{\mathrm{mf}}(\mathbf{k},-\omega_{n}) \right]^{2}
\nonumber \\
& \simeq & \int \!\!\! \frac{d \mathbf{k}}{(2 \pi)^{3}} \, \frac{1}{4 \, E(\mathbf{k})^{2}} \simeq \frac{m \, k_{F}}{8 \, \pi \, \Delta} = \frac{m^{2} \, \xi_{0}}{8} 
\label{form-factors-BEC-limit}
\end{eqnarray}

\noindent
where $\xi_{0} = \frac{k_{F}}{\pi m \Delta}$ is the Pippard coherence length characteristic of BCS superconductivity \cite{FW-1971}.
On the other hand, in Eq.~(\ref{dominant-term-BCS-limit}) the sum over the bosonic Matsubara frequencies of the $11$-component of the particle-particle ladder can conveniently be dealt via the spectral representation \cite{PPS-2004} 
\begin{equation}
\Gamma_{11}(\mathbf{Q},\Omega_{\nu}) = - \! \int_{-\infty}^{+\infty} \! \frac{d \omega}{\pi} \, \frac{\mathrm{Im} \, \Gamma_{11}^{R}(\mathbf{Q},\omega)}{i\Omega_{\nu} - \omega} \, ,
\label{spectral-representation}
\end{equation}
where the spectral function $\Gamma_{11}^{R}(\mathbf{Q},\omega)$ with real frequency $\omega$ is obtained from $\Gamma_{11}(\mathbf{Q},\Omega_{\nu})$ given by Eqs.~(\ref{full-particle-particle-ladder})-(\ref{B}) 
with the replacement $i\Omega_{\nu} \rightarrow \omega + i \eta$.
At zero temperature, in the small-$Q$ limit the spectral function in Eq.~(\ref{spectral-representation}) is expected to take the form
\begin{eqnarray}
\mathrm{Im} \, \Gamma_{11}^{R}(\mathbf{Q},\omega) & = & \alpha_{+}(\mathbf{Q}) \, \delta(\omega - E_{AB}(\mathbf{Q}))
\nonumber \\
& + & \alpha_{-}(\mathbf{Q}) \, \delta(\omega + E_{AB}(\mathbf{Q})) \, ,
\label{spectral-function}
\end{eqnarray}
where $E_{AB}(\mathbf{Q}) = s \, Q$ is the dispersion relation of the Anderson-Bogoliubov mode with sound velocity $c$.
In this limit, one then gets:
\begin{equation}
\frac{1}{\beta} \sum_{\nu} e^{i \Omega_{\nu} \eta} \, \Gamma_{11}(\mathbf{Q},\Omega_{\nu}) = - \, \frac{\alpha_{-}(\mathbf{Q})}{\pi} \, .
\label{relevant-Matsubara-sum}
\end{equation}
In turn, the amplitude $\alpha_{-}(\mathbf{Q})$ can be obtained by expanding the expressions (\ref{full-particle-particle-ladder})-(\ref{B}) (where $i\Omega_{\nu} \rightarrow \omega + i \eta$) for small $Q$ and $\omega$,
yielding at the relevant order
\begin{equation}
\Gamma_{11}^{R}(\mathbf{Q},\omega) \simeq \frac{a_{0}}{F \, \left[ s^{2} \, Q^{2} - (\omega +  i \eta)^{2} \right]} \, .
\label{small-Q_and_omega-expansion}
\end{equation} 
From this expression we get
\begin{equation}
\mathrm{Im} \, \Gamma_{11}^{R}(\mathbf{Q},\omega) \simeq \frac{a_{0}} {2 F s Q} \, \mathrm{sgn}(\omega) \, \left[ \delta(\omega - s Q) + \delta(\omega + s Q) \right] \, ,
\label{small-Q_and_omega-expansion-imaginary_part}
\end{equation}
such that
\begin{equation}
\alpha_{-}(\mathbf{Q}) = - \frac{a_{0} \pi}{2 F s} \, \frac{1}{Q}
\label{expression-alpha}
\end{equation}
where $F s = \sqrt{2 a_{0} (a_{2} - b_{2}) [2 a_{0} (b_{3} - a_{3}) + a_{1}^{2}]}$.

The expressions of the coefficients $(a_{0}, a_{1}, a_{2}, a_{3}, b_{2}, b_{3})$ needed in Eqs.~(\ref{small-Q_and_omega-expansion-imaginary_part}) and (\ref{expression-alpha}) are provided in Ref.~\cite{MPS-1998}, 
where they are calculated analytically in terms of elliptic integrals.
In particular, in the BCS limit 
\begin{equation}
\frac{ \left( a_{1} \right)^{2} }{ 2 a_{0} (b_{3} - a_{3}) }  \simeq \frac{1}{4} \, \left[ \frac{ \ln (8 x_{0}) }{ x_{0} } \right]^{2}  \ll  1 
\label{small-fraction}
\end{equation}
since $x_{0} = \mu / \Delta \simeq E_{F} / \Delta \gg 1$ in this limit.
Accordingly, in the expression (\ref{expression-alpha}) for $\alpha_{-}(\mathbf{Q})$ one approximates
\begin{equation}
\frac{a_{0}}{2 F s} \simeq \frac{1}{4 \sqrt{(a_{2} - b_{2}) (b_{3} - a_{3})}} \simeq \frac{ \sqrt{3} \, \pi^{2} \Delta }{m \, x_{0}} \simeq \frac{2 \sqrt{3}}{m^{2} \xi_{0}^{2}} \, ,
\label{coefficient-alpha-approximate}
\end{equation}
which has also been expressed in terms of the Pippard coherence length $\xi_{0}$.

In conclusion, the results (\ref{form-factors-BEC-limit}), (\ref{relevant-Matsubara-sum}), (\ref{expression-alpha}), and (\ref{coefficient-alpha-approximate}) can be entered in the expression (\ref{dominant-term-BCS-limit}), yielding for the
leading asymptotic spatial behavior of $\delta h(R)$: 
\begin{equation}
\delta h(R) \simeq \frac{\sqrt{3}}{4 \, \xi_{0}} \, \int \!\!\! \frac{d \mathbf{Q}}{(2 \pi)^{3}} \, \frac{ e^{i \mathbf{Q} \cdot \mathbf{R}}}{ |\mathbf{Q}| } =  \frac{\sqrt{3}}{8 \pi^{2} \, \xi_{0}} \, \frac{1}{R^{2}} \, .
\label{BCS-asymptotic-spatial-behavior-delta_h}
\end{equation}

As a final comment, it might be interesting to compare the result (\ref{BCS-asymptotic-spatial-behavior-delta_h}) obtained in the BCS limit at zero temperature with the corresponding result (\ref{bosonic-normal-propagator-equal-time-asymptotic}) obtained in the BEC limit.
This comparison can conveniently be made in terms of the phase coherence length $\xi_{\mathrm{phase}}$ discussed in Ref.~\cite{PS-1996} at zero temperature, which, at the level of the pairing fluctuations here considered, reduces to the healing length $\xi_{B}$
of the Bogoliubov theory in the BEC limit and to $(\pi/6) \xi_{0}$ in the BCS limit (see also Ref.~\cite{PS-1994}).
The BCS result (\ref{BCS-asymptotic-spatial-behavior-delta_h}) can thus be cast in the form
\begin{equation}
\delta h(R) \simeq  \left( \frac{\pi}{2 \, \sqrt{3}} \right) \, \frac{1}{8 \pi^{2} \xi_{\mathrm{phase}} \, R^{2} } \, ,
\label{BCS-asymptotic-spatial-behavior-delta_h-phase}
\end{equation}
which differs from the BEC result (\ref{bosonic-normal-propagator-equal-time-asymptotic}) by the factor $\frac{\pi}{2 \, \sqrt{3}} \simeq 0.91$.

The results (\ref{bosonic-normal-propagator-equal-time-asymptotic}) for the BEC limit and (\ref{BCS-asymptotic-spatial-behavior-delta_h-phase}) for the BCS limit have suggested us to introduce 
the length $\xi_{\mathrm{odlro}}$ associated with the asymptotic spatial behavior of $\delta h(R)$ in the context of the ODLRO, which at zero temperature 
evolves from $\xi_{\mathrm{phase}}$ in the BEC limit to $\frac{2 \, \sqrt{3}}{\pi}\xi_{\mathrm{phase}} \simeq 1.1 \, \xi_{\mathrm{phase}}$ in the BCS limit
(cf. the inset of \color{red}Fig.~\ref{Figure-7}\color{black}).
The values of $\xi_{\mathrm{odlro}}$ identified from the asymptotic spatial behavior of $\delta h(R)$ were then obtained numerically throughout the BCS-BEC crossover Sec.~\ref{subsec:results_fluctuations}, 
not only at zero temperature but also where it was possible for temperatures close to $T_{c}$.
 
	


\begin{thebibliography}{99}
\bibitem{footnote-4} Recall in this context that the Bogoliubov approximation for a degenerate almost ideal Bose gas at low temperature is expected to hold for $Q$ up to a maximum of the order $\tilde{\xi}_{B}^{-1}$ \cite{LP-2006}.
                                However, the Bogoliubov approximation yields also the correct $C/Q^{4}$ behavior for the quasi-particle occupation number $v_{B}(Q)^{2}$ when $Q \, \xi_{B} \gg 1$, 
                                with the value $( 2 \, \tilde{\xi}_{B}^{4})^{-1} = (8 \pi a_{B} n_{B})^{2}/2$ for the Tan contact $C$.
\end{thebibliography}
\end{document}